\documentclass[preprint,12pt]{elsarticle}

\usepackage{graphicx}
\usepackage{placeins}
\usepackage{subcaption}

\usepackage{amssymb}
\usepackage{amsmath}
\usepackage{amsthm}
\usepackage{amsfonts}

\usepackage{hyperref}
\usepackage[usenames,dvipsnames]{xcolor}

\makeatletter
\def\ps@pprintTitle{%
     \let\@oddhead\@empty
     \let\@evenhead\@empty
     \def\@oddfoot{\footnotesize\itshape
     {\copyright 2017. This manuscript version is made available under the
      \href{https://creativecommons.org/licenses/by-nc-nd/4.0/}{CC-BY-NC-ND 4.0 license}.}
      \hfill}
     \let\@evenfoot\@oddfoot}

\usepackage{color}

\usepackage{changes}
\definechangesauthor[name={V}, color=Fuchsia]{vl}
\definechangesauthor[name={E}, color=red]{er}
\newcommand\revE[2]{{#1}{#2}}

\newtheorem{remark}{Remark}[section]

\def\eq#1{(\ref{#1})}

\def\bmi#1{\textbf{\textit{#1}}}
\def\ol#1{\overline{#1}}
\def\dvg{\mbox{\rm div}}

\def\sigmabf{{\mbox{\boldmath$\sigma$\unboldmath}}}
\def\omegabf{{\mbox{\boldmath$\omega$\unboldmath}}}
\def\xibf{{\mbox{\boldmath$\xi$\unboldmath}}}

\def\veps{\varepsilon}
\def\vrho{\varrho}
\def\vphi{\varphi}
\def\pd{\partial}
\def\Om{\Omega}
\def\RR{{\mathbb{R}}}
\def\ZZ{{\mathbb{Z}}}
\def\intY{\sim \kern-1.2em \int}
\def\intYs{\smallsim \kern-.75em \int}
\newcommand{\smallsim}{\ensuremath \raisebox{.15em}{{$\scriptstyle\sim$}}}

\def\dV{\mathrm{\,dV}}  %
\def\dVx{\mathrm{\,dV}_{\kern-0.1em x}}  %
\def\dVxy{\mathrm{\,dV}_{\kern-0.1em xy}}  %
\def\dY{\mathrm{\,dV}_{\kern-0.1em y}}  %
\def\dS{\mathrm{\,dS}}  %
\def\dSx{\mathrm{\,dS}_{x}}  %
\def\dSy{\mathrm{\,dS}_{y}}  %
\def\mx{{[m]}}
\def\mxe{{[m*]}}
\def\cx{{[c]}}
\def\rhs{{r.h.s.{~}}}
\def\mic{{\rm{mic}}}
\def\eeb#1{\eb({#1})}

\def\eeby#1{\eb_y({#1})}
\def\eebx#1{\eb_x({#1})}
\def\ext{{\rm{ext}}}

\def\aYms#1#2{a_{Y}^{m*} \left ({#1},\,{#2}\right )}

\def\gYm#1#2{g_{Y}^m \left ({#1},\,{#2}\right )}
\def\dYm#1#2{d_{Y}^m \left ({#1},\,{#2}\right )}

\newcommand\wrt{{w.r.t.{~}}}

\newcommand\ie{{\it{i.e.~}}}
\newcommand\eg{{\it{e.g.~}}}
\newcommand\cf{{{cf.~}}}
\newcommand\wtilde[1]{\widetilde{#1}}
\newcommand\ul[1]{\underline{#1}}

\def\Pcal{\mathcal{P}}

\def\Vcal{\mathcal{V}}
\def\Wcal{\mathcal{W}}
\def\Ycal{\mathcal{Y}}
\def\Zcal{\mathcal{Z}}

\def\Ucalbf{{\mbox{\boldmath$\mathcal{U}$\unboldmath}}}

\newcommand\cwto{\rightharpoonup}

\newcommand\Grxy[2]{\left(\nabla_x{#1} + \nabla_y{#2}\right)}
\newcommand\GrxyS[2]{\left(\nabla_x^S{#1} + \nabla_y^S{#2}\right)}

\newcommand{\integer}[1]{{\rm{int}}\left\{{#1}\right\}}
\newcommand{\dist}[2]{{\rm{dist}}\left({#1},{#2}\right)}
\newcommand{\jump}[1]{\left[{#1}\right]}

\newcommand\nrm[2]{\left\| {#1}\right\|_{#2}}  
\newcommand{\Tuf}[1]{{\mathcal{T}}_\veps{\left ({#1}\right )}}

\def\ab{{\bmi{a}}}
\def\bb{{\bmi{b}}}
\def\ub{{\bmi{u}}}
\def\vb{{\bmi{v}}}
\def\wb{{\bmi{w}}}
\def\fb{{\bmi{f}}}
\def\hb{{\bmi{h}}}
\def\nb{{\bmi{n}}}
\def\db{{\bmi{d}}}
\def\eb{{\bmi{e}}}
\def\gb{{\bmi{g}}}
\def\Bb{{\bmi{B}}}
\def\Db{{\bmi{D}}}

\def\Ub{{\bmi{U}}}
\def\Wb{{\bmi{W}}}
\def\Gb{{\bmi{G}}}
\def\Kb{{\bmi{K}}}
\def\Sb{{\bmi{S}}}
\def\kb{{\bmi{k}}}
\def\Fb{{\bmi{F}}}

\def\Pibf{{\mbox{\boldmath$\Pi$\unboldmath}}}
\def\Hdb{{\bf{H}}^1}
\def\Hpdb{{\bf{H}}_\#^1}
\def\Hdb{{\bf{H}}^1}
\def\Hb{{\bf{H}}}

\def\Lb{{\bf{L}}}

\def\Dop{{{\rm I} \kern-0.2em{\rm D}}}
\def\Aop{{{\rm A} \kern-0.6em{\rm A}}}%
\def\coefF{\Fb^H}
\def\coefaF{\acute \Fb^H}
\def\coefG{\ul{\Gb}^H}
\def\coefaG{\ul{\acute \Gb}^H}
\def\coefGU{\ul{\Gb}^U}
\def\parg{\ul{\gb}}

\newcommand\norm[1]{\left\lVert#1\right\rVert}

\begin{document}

\begin{frontmatter}

\title{Homogenization of the fluid-saturated piezoelectric porous media\tnoteref{preprint}}

\tnotetext[postprint]{Preprint submitted to Elsevier.}

\author[NTIS]{E.~Rohan\corref{cor1}}
\ead{rohan@kme.zcu.cz}
\cortext[cor1]{Coresponding author.}
\author[NTIS]{V.~Luke\v{s}}
\ead{vlukes@kme.zcu.cz}

\address[NTIS]{Department of Mechanics \& 
  NTIS New Technologies for Information Society, Faculty of Applied Sciences, University of West Bohemia in Pilsen, \\
Univerzitn\'\i~22, 30100 Plze\v{n}, Czech Republic}

\begin{abstract}
The paper is devoted to the homogenization of porous piezoelectric materials saturated by electrically inert fluid.
The solid part of a representative volume element consists of the piezoelectric skeleton with embedded conductors. The pore fluid in the periodic structure can constitute a single connected domain, or an array of inclusions. Also the conducting parts are represented by several mutually separated connected domains, or by inclusions. Two of four possible arrangements are considered for upscaling by the homogenization method. The macroscopic model of the first type involves coefficients responsible for interactions between the electric field and the pore pressure, or the pore volume. For the second type, the electrodes can be used for controlling the electric field at the pore level, so that the deformation and the pore volume can be influenced locally. Effective constitutive coefficients are computed using characteristic responses of the microstructure. The two-scale modelling procedure is implemented numerically using the finite element method. The macroscopic strain and electric fields are used to reconstruct the corresponding local responses at the pore level. For validation of the models, these are compared with results obtained by direct numerical simulations of the heterogeneous structure; a good agreement is demonstrated, showing relevance of the two-scale numerical modelling approach.
\end{abstract}


\end{frontmatter}



\section{Introduction}\label{sec-intro}

The piezoelectric effects which couple the mechanical deformation and the
electrical field have been studied since the middle of eighteen century, however
the piezoelectric materials became widespread during the World War~I due to
their use in resonators for detecting the acoustic sources produced by
submarines using echolocation. After the World War~II, apart of quartz, new
types of the piezoelectric materials, such as barium titanite (\emph{BaTiO$_3$})
and other synthesizes piezoceramic materials were developed with their
dielectric constants much higher than those found in natural piezoelectric
materials, such as quartz and some other minerals, or bone.    Since then, the
piezoelectric materials have found wast applications in electronics,
mechatronics, and micro-system technology, being extensively used in the design
of transducers, sensors and energy harvesters. Smart structures, such as
microelectromechanical systems (MEMS) based on these materials allow for
intelligent self-monitoring and self-control capabilities. Nowadays the
piezoelectric sensor-actuator systems can be distributed continuously, being
attached to the surface of other structural parts.  Such an arrangement can be
used \eg in the aerospace industry to control vibrations, or acoustic radiation
of thin flexible structures.

For modelling periodically heterogeneous media with piezoelectric components,
classical upscaling techniques has been employed. Besides the micromechanics
approaches including the Mori-Tanaka and self consistent upscaling schemes,
\cite{Ayuso-etal-IJSS2017-pz-homog}, the classical periodic homogenization based
on the formal two-scale asymptotic expansion method
\cite{Sanchez1980Book,Cioranescu1999book}, or on the two-scale convergence
\cite{Allaire1992} and the periodic unfolding method \cite{Cioranescu2008a} has
been used. Recently, the homogenization of thermoelectric materials was treated
in \cite{Fantoni-IJSS2017-thermo-pz}. Homogenization of the periodic  composites
consisting of piezoelectric matrix and elastic anisotropic inclusions accounting
for bone cells was described in \cite{Miara-piezo}; therein it has been
suggested to exploit the piezoelectric effect in the design of a new type of
bio-materials which should assist in bone healing and regeneration. Such
possible application for piezoelectric materials in biomedical engineering is
motivated by the electrochemical processes in biological tissues, which are
coupled tightly with periodic mechanical loading assisted by the electric field.
Performance of the tissue regeneration and remodelling may be enhanced by
activated bio-piezo porous implants which can accelerate these processes
undergoing at the microscopic level and related to the electro-mechanical
transduction, cf. \cite{Sikavitsas-etal-2001,Wiesmann-etal-2001}. The Suquet
method of homogenization has been used to obtain analytic models of particulate
and fibrous piezoelectric composites \cite{Iyer-IJSS-2014}. Apart of the
homogenization of heterogeneous piezoelectric media, an asymptotic analysis has
been applied to derive higher order models of piezoelectric rods and beams,
starting from the 3D piezoelectricity problem \cite{Viano-pz-beams-IJSS2016}.

In \cite{Rohan-piezo-sa}, the shape sensitivity formulae were derived for a
class of 2D microstructures comprising one piezoelectric and one arbitrary
elastic material, whereby the shape of the interface between the two materials
was parameterized. As a challenge for the material design, the numerical tests
have shown how a suitable geometry of the interface can amplify some of the
homogenized coefficients, namely the third-order tensors associated with the
electromechanical coupling. Sensitivity of the effective medium properties to
the microstucture properties were also reported in \cite{Koutsawa-sa-2010}.

Besides the periodic homogenization, in \cite{Ayuso-etal-IJSS2017-pz-homog}, the
Mori-Tanaka and the self-consistent schemes were used for upscaling the drained
porous piezoelectric materials. Concerning the fluid saturated porous
piezoelectric media, the asymptotic method has been applied in
\cite{Telega-Wojnar-CRAS2b-2000} to derive macroscopic constitutive laws
accounting for the fluid-structure interaction at the pore level, whereby a
simplified model of electrolytes was considered. In the context of the bone
tissue biomechanics, the macroscopic influence of piezoelectric effects
observable in dried bone  was studied in \cite{Lemaire-etal2011-bone-pz} using
the homogenization approach.

Propagation of electroacoustic waves in an reinforced piezoelectric medium was
treated in \cite{Levin-IJSS2002-pz-waves}. The low frequency acoustic wave
propagation in the porous piezoelectric materials  has been subject of several
works
\cite{Vashishth-wave-pz-2009,Sharma-PRSA2010-waves-pz,VG-waves-pz-JASA2011,Besombes-etal-JASA1990}.
In these papers, the modelling is based on the Biot theory of porous media
elaborated within the phenomenological approach, therefore, influences of
specific microstructures on the wave dispersion have not been studied yet.

This paper is focused on the derivation of the effective material coefficients
of the fluid-saturated porous media with the piezoelectric skeleton using the
homogenization framework. A related topic was treated recently in
\cite{Iyer-IJSS-2014}, where a special type of piezoelectric anisotropic
composite materials was studied using numerical and analytical methods. Although
the porosity influence was examined and the figures of merit related to the
hydrostatic strain coefficient were also investigated, we pursue another
homogenization approach which is based on the periodic homogenization of the
static fluid-structure interaction, as reported in
\cite{rohan-etal-CMAT2015-porel} in the context of the hierarchical porous
poroelastic media, cf. \cite{Rohan-AMC}. We assume a quasistatic loading, such
that inertia and viscosity related effects can be neglected. As the consequence,
in any connected porosity, a unique pressure is established which satisfies the
equilibrium. Using the homogenization of the fluid-structure interaction problem
at the microscopic scale, we obtain macroscopic models of the upscaled
piezo-poroelastic medium for different periodic microstructures; one connected
porosity, or an array of fluid filled inclusions is combined with piezoelectric
skeleton which can contain mutually separated conductors (metallic parts). We
consider two different situations: 1) the conductors are distributed as a
periodic arrays of mutually separated inclusions, or 2) the conducting pars
constitute two or more electrodes such that each of these electrodes presents a
connected porous structure. In the second case, different electric potential is
prescribed to different electrodes, so that electric fields induced in the
microstructure can be controlled.

The paper is organized as follows. In Section~\ref{sec-mic-pb}, different
microscopic configurations of the periodic porous piezoelectric medium are
defined and the model equations are introduced, yielding the weak formulation.
The homogenization of the static fluid-structure interaction is reported in
Sections~\ref{sec-CFD*} and \ref{sec-DFC*}, for the two above mentioned designs
of the conducting parts. In both these sections, the local problems for the
characteristic responses of the representative period cell are derived and the
homogenized effective material coefficients are obtained. These are involved in
the macroscopic equations governing behaviour of the upscaled poro-piezoelectric
medium. Using the characteristic responses and the macroscopic fields, the
displacement, pressure and electric fields can be reconstructed at the
microscopic  level, as reported in Section~\ref{sec-reconstruct}. Finally, in
Section~\ref{sec-num-ex}, we present numerical illustration of the derived
macroscopic models. For validation of these models, direct numerical simulation
of the heterogeneous media are compared with the responses computed using the
homogenized problems. Some technical supporting material is explained
in the Appendix.

\paragraph{Some basic notations} In the paper, the mathematical models are
formulated in a Cartesian framework of reference $\mathcal{R}(\text{O};
\eb_1,\eb_2,\eb_3)$ where $O$ is the origin of the space and
$(\eb_1,\eb_2,\eb_3)$ is a orthonormal basis for this space. The coordinates of
a point $M$ are specified by $x=(x_1, x_2, x_3)$ in $\mathcal{R}$. The boldface
notation for vectors, $\ab = (a_i)$, and for tensors,  $\bb = (b_{ij})$,
is used. The following special notation is used for the electric field $\vec E$, and the electric displacement vector $\vec D$. Furthermore, a special notation is introduced for the 3rd order tensors associated with piezoelectric coupling, $\coefG, \parg$.
 The gradient and divergence operators are respectively
denoted by $\nabla$ and $\nabla \cdot$. When these operators have a subscript
which is space variable, it is for indicating that the operator acts relatively
at this space variable, for instance $\nabla_x = (\pd_i^x)$. The small strain
tensor is denoted by $\eeb{\ub^\veps} = (\nabla\ub^\veps +
(\nabla\ub^\veps)^T)/2$. The symbol dot `$\cdot$' denotes the scalar product
between two vectors and the symbol colon `$:$' stands for scalar (inner) product
of two second-order tensors.
Throughout the paper, $x$ denotes the global (``macroscopic'') coordinates, while the ``local''
coordinates $y$ describe positions within the representative unit cell
$Y\subset\RR^3$ where $\RR$ is the set of real numbers.
\revE{By $\dV$ (or $\dVx$) and  $\dY$  we denote the elementary volume elements associated with coordinates $x$ and $y$, respectively, while $\dVxy$ is  the elementary volume in a cross-product domain $\Om\times Y$. Accordingly, elementary surfaces are designated by $\dS$, $\dSx$ and $\dSy$.}{}

 By $\intYs_{Y_d} = |Y|^{-1}\int_{Y_d}$ with $Y_d\subset \ol{Y}$ we denote the
 local average. 
 The Lebesgue spaces of
 2nd-power integrable functions on a domain $D$ is denoted by $L^2(D)$, the
 Sobolev space $\Wb^{1,2}(D)$ of the square integrable vector-valued functions
 on $D$ including the 1st order generalized derivative, is abbreviated by
 $\Hdb(D)$. The unit normal vector outward to domain $D_s$ is denoted by
 $\nb^{[s]}$.

\section{Microscopic model of porous piezoelectric media}\label{sec-mic-pb}

There are typically two characteristic lengths: $\ell$ describes the
heterogeneity size and $L$ is the relevant macroscopic size. The ration $\veps =
\ell/L$ is called the scale parameter. As usually, we consider material
properties of the heterogeneous medium oscillating with period $\ell$ relative
to the spacial position. The asymptotic method of homogenization is based on the
asymptotic analysis of the mathematical model for $\veps\rightarrow 0$.

\subsection{Periodic microstructure}\label{sec-period}

\begin{figure}
\centering
\includegraphics[width=0.65\linewidth]{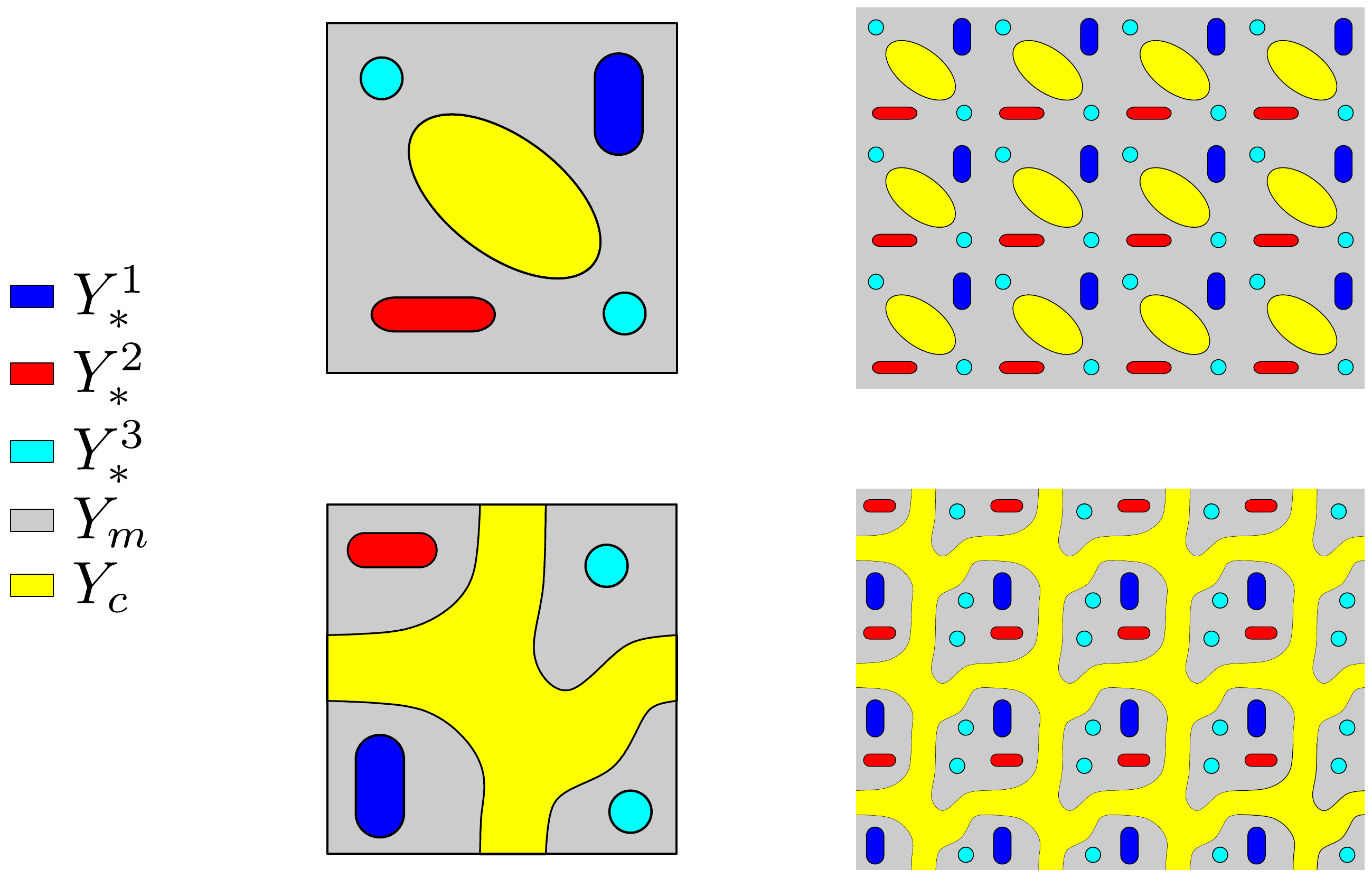}
\caption{The scheme of the representative periodic cell decomposition and the
generated periodic structure. Two configuration of the porous microstructure
with conducting parts $Y_*^k$ embedded in the solid piezoelectric skeleton
$Y_m$.}\label{fig-CFDF}
\end{figure}

The medium is generated by copies of the representative volume element (RVE)
$Z^\veps\subset \RR^3$ as a periodic lattice, so that $\veps a_k$ is the lattice
period in the $k$-the coordinate direction. For the ``real size'' RVE, we
introduce its rescaled copy $Y = \veps^{-1} Z^\veps$ which is called the
rescaled elementary periodic cell $Y$ defined by $Y =
\prod_{i=k}^3]-a_k/2,a_k/2[$; typically $|Y| = 1$, see Fig.~\ref{fig-CFDF}. In
this paper we consider $a_k = 1$ without loss of generality. For any given
$\veps >0$ we define mesoscopic (zoomed) coordinates $y=(y_k) \in Y$ which for a
given ``macroscopic'' position $x$ are given by the localization function
$\Ycal:x\mapsto y$ defined by: $y_k = \Ycal_k(x) = \left(x_k - \veps
a_k\integer{x_k/(\veps a_k)}\right)/\veps$ for $k=1,2,3$, where $\integer{z}$
denotes the integer part of $z$.

\subsection{Porous piezoelectric solid saturated by static fluid}

We consider a quasi-static loading of a piezoelectric skeleton interacting with
a viscous fluid saturating pores in the skeleton.

The piezo-poroelastic medium occupies an open bounded domain $\Om \subset \RR^3$
with Lipschitz boundary $\pd \Om$.
The following decomposition of $\Om $ into the piezoelectric matrix,
$\Om_m^\veps$, elastic conductive inclusions, $\Om_*^\veps$, and fluid-saturated
channel parts, $\Om_c^\veps$, is considered:
\begin{equation}\label{eq-pzm1}
\begin{split}
\Om & = \Om_c^\veps \cup \Om_m^\veps \cup \Om_*^\veps\;,\quad \Om_c^\veps \cap \Om_m^\veps \cap \Om_*^\veps = \emptyset\;,\\
\mbox{ where } \Om_*^\veps & = \bigcup_k \Om_*^{k,\veps}\;.
\end{split}
\end{equation}
By $\Gamma_c^\veps$ we denote the solid-fluid interface, $\Gamma_c^\veps  =
\ol{\Om_m^\veps \cup \Om_*^\veps} \cap \ol{\Om_c^\veps}$. The interface between
the piezoelectric matrix and the conductors $\Gamma_*^{\veps}$ consists of its
subparts $\Gamma_*^{k,\veps}$ introduced, as follows:
\begin{equation}\label{eq-pzm1b}
\Gamma_*^{\veps} = \pd \Om_*^{\veps} \cap \ol{\Om_m^\veps}\quad \mbox{ and }\quad \Gamma_*^{k,\veps} = \pd \Om_*^{k,\veps} \cap \ol{\Om_m^\veps}\;.
\end{equation}
Further we denote by $\pd_\ext\Om_m^\veps = \ol{\Om_m^\veps} \cap \pd\Om$ the exterior boundaries of $\Om_m^\veps$.
In analogy, we define $\pd_\ext\Om_*^\veps$ and $\pd_\ext\Om_c^\veps$ as the exterior parts on the boundaries of the conductive solid $\Om_*^\veps$ and the fluid $\Om_c^\veps$, respectively.  We assume that $\Om_m^\veps$ is connected domain, however
$\Om_c^\veps$ may consists of disconnected inclusions; the latter option will be considered as a special case. The conductive material is distributed as a connected phase for each index $k$, such that $\Om_*^{k,\veps}$ is connected.

As we often refer to the solid part consisting of the piezoelectric matrix and
the conductor part, we introduce $\Om_{m*} = \Om_m^\veps \cup \Om_*^\veps \cup
\Gamma_*^{\veps}$, recalling the interface $\Gamma_*^{\veps}$ between the two
parts is defined in \eq{eq-pzm1b}.
The boundary conditions are prescribed on  the external boundary
$\pd_\ext\Om_{m*} = \pd\Om_{m*}\setminus \Gamma_*^{\veps}$ involving both the
solid phases; the following two splits are defined, $\pd_\ext\Om_{m*}^\veps =
\Gamma_{\ub}^\veps \cup \Gamma_\sigma^\veps$ and $\pd_\ext\Om_{m*}^\veps =
\Gamma_{\vphi}^\veps \cup \Gamma_{\vec D}^\veps$ such that
\begin{equation}\label{eq-pzm1a}
\Gamma_\sigma^\veps = \pd_\ext\Om_{m*}^\veps\setminus\Gamma_{\ub}^\veps\;\quad\mbox{and}\quad
\Gamma_{\vec D}^\veps = \pd_\ext\Om_{m*}^\veps\setminus\Gamma_{\vphi}^\veps\;.
\end{equation}

To respect spatial fluctuations of the material parameters, by virtue of the
scale parameter introduced above, all material coefficients and unknown
functions involved in the mathematical model which depend on the scale will be
labeled by superscript $\veps$. In the piezoelectric solid, the Cauchy stress
tensor $\sigmabf^\veps$ and the electric displacement $\vec D^\veps$ depend on
the strain tensor $\eeb{\ub^\veps} = (\nabla\ub^\veps + (\nabla\ub^\veps)^T)/2$
defined in terms of the displacement field $\ub^\veps = (u_i^\veps)$, and on the
electric field $\vec E{\vphi^\veps} = \nabla \vphi^\veps$ defined in terms of the
electric potential $\vphi^\veps$. The following constitutive equations
characterize the piezoelectric solid in $\Om_m^\veps$,
\begin{equation}\label{eq-1}
\begin{split}
\sigma_{ij}^\veps(\ub^\veps,\vphi^\veps) & = A_{ijkl}^\veps e_{kl}^\veps(\ub^\veps) - g_{kij}^\veps E_k^\veps(\vphi^\veps)\;,\\
D_k^\veps(\ub^\veps,\vphi^\veps) & =  g_{kij}^\veps e_{ij}^\veps(\ub^\veps) + d_{kl}^\veps E_l^\veps(\vphi^\veps)\;,
\end{split}
\end{equation}
where $\Aop^\veps=(A_{ijkl}^\veps)$ is the elasticity fourth-order symmetric
positive definite tensor of the solid, where $A_{ijkl} = A_{klij} =
A_{jilk}$,the deformation is coupled with the electric field through the 3rd
order tensor $\parg^\veps = (g_{kij}^\veps)$, $g_{kij}^\veps = g_{kji}^\veps$ and
$\db = (d_{kl})$ is the permitivity tensor. The conductive solid is described by
its elasticity $\Aop^\veps$ only. The permitivity in $\Om_*^{k,\veps}$ is
infinitely large, so that we assume $\vphi^\veps = \bar\vphi^k$.

In this paper we shall use a compact (global) notation, such that \eq{eq-1} can
be written (we drop the superscript $^\veps$ for the moment), $\sigmabf =
\Aop\eeb{\ub} - \parg^T\cdot\vec E(\vphi)$, and $\vec D = \parg:\eeb{\ub} + \db\vec
E$.

\subsection{Discussion of possible configurations and the problem formulations}

In the context of the steady state -- thermodynamic equilibrium under no fluid
flow, two situations will be considered, see Fig.~\ref{fig-CFDF}:
\begin{itemize}

\item  Case CF -- connected porosity (fluid canals) $\Om_c^\veps$: in the
thermodynamic equilibrium, the pore fluid pressure is constant (denoted by $\bar
p$);

\item Case DF -- disconnected fluid inclusions: each $\Om_c^{k,\veps}$ is formed
by inclusions with diameter $\approx\veps$;  such an arrangement admits
differences in the fluid pressure associated with neighbouring inclusions.

\end{itemize}
The first case can be modified for a quasi-static flow in the channels with
neglected inertia effects and assuming moderate pressure gradients in the flow,
as considered in \cite{Rohan-AMC}, cf. \cite{RSW-ComGeo2013}.

Concerning the conductor part, two different configurations are assumed:
\begin{itemize}

\item Case D* -- disconnected conductor inclusions: each $\Om_*^{k,\veps}$,
$k=1,2,\dots,\bar k^\veps$ is formed by inclusions with diameter $\approx\veps$,
thus, the $k$-th inclusion $\Om_*^{k,\veps}$ is contained in a copy of
$Z^\veps$. Obviously, the number of such inclusions depends on $\veps$;

\item Case C* -- connected conductor fibres: each $\Om_*^{k,\veps}$ is a
connected domain, however, they are mutually disconnected. We consider a finite
number $k^*$ of such conductors.

\end{itemize}
Each of the two cases, D* and C*, needs a special treatment with respect to the
homogenization $\veps\rightarrow 0$ which is the subject of
Section~\ref{sec-CFD*} and \ref{sec-DFC*}, respectively.


\subsection{Problem formulations}

With reference to the notation related to the domain decomposition and the
associated parts of the interfaces and the external boundary, as established
above, we introduce  differential equations, the interface and the boundary
conditions governing static behaviour, or steady state behaviour of the porous
fluid saturated piezoelectric medium. The boundary value problems (BVPs) are
constituted by the following equations involving the $\ub^\veps,\vphi^\veps$ and
$p^\veps$:
\begin{itemize}
\item Equilibrium of the stress and electric displacements,
\begin{equation}\label{eq-2}
\begin{split}
-\nabla\cdot \sigmabf^\veps(\ub^\veps,\vphi^\veps) & = \fb^\veps\;, \quad \mbox{ in }  \Om_{m*}^\veps\;, \\
-\nabla\cdot \vec D^\veps(\ub^\veps,\vphi^\veps) & = q_E^\veps\;,\quad \mbox{ in }  \Om_m^\veps\;,
\end{split}
\end{equation}
where $\fb^\veps$ is the volume-force and $q_E^\veps$ is the volume electric
charge;
\item The fluid mass conservation is presented for the two cases, CF and DF.

Inclusions (DF): Let $\Om_{c}^{k,\veps} \subset \Om_c^\veps$ be the $k$-th
inclusion, the mass conservation yields
\begin{equation}\label{eq-3a}
\int_{\pd \Om_c^{k,\veps}} {\ub^\veps}\cdot \nb^\cx \dS + \gamma   p^{k,\veps} |\Om_c^{k,\veps}|= 0\;, \quad \forall k \in \{1,\dots,\bar k^\veps\}\;,
\end{equation}
where $\gamma $ is the fluid compressibility.

Canals (CF):  In the steady state, the fluid pressure gradient vanishes,
so that one pressure value is attained in the whole connected porosity
$\Om_c^\veps$. The fluid volume $-J^\veps$ injected through the boundary
$\pd_\ext\Om_c^\veps$ into the porosity is compensated by the pore inflation and
by the fluid compression, so that \eq{eq-3a} is replaced by:
\begin{equation}\label{eq-3b}
\int_{\pd \Om_c^\veps} {\ub^\veps}\cdot \nb^\cx \dS + \gamma   p^\veps |\Om_c^\veps|= -J^\veps\;.
\end{equation}

In general, the connected porosity admits flow at the global level of the
connected porosity $\Om_c^\veps$. In this case, \eq{eq-3a} is replaced by
\begin{equation}\label{eq-3c}
\int_{\pd \Om_c^\veps} q^\veps\left( {\dot\ub^\veps}\cdot \nb^\cx + \gamma \dot p^\veps  + \nabla\cdot \wb^\veps\right)\dS = 0\;,
\end{equation}
where $\wb^\veps$ is the seepage velocity governed by the Stokes flow and the
dot means the time derivative. However, in this paper we shall consider
stationary problems only, thus, governed by equations \eq{eq-3a} or \eq{eq-3b}.

\item Interface conditions:
\begin{equation}\label{eq-4a}
\begin{split}
\nb \cdot \sigmabf^\veps  & = - p^\veps \nb\;,\quad \mbox{ on } \Gamma_c^\veps\;, \\
\nb^\mx \cdot \vec D^\veps  & = \varrho_E^\veps\;,\quad \mbox{ on } \Gamma_{mc}^\veps\;,\\
\nb \cdot \jump{\sigmabf^\veps} & = 0\;,\quad \mbox{ on } \Gamma_*^\veps\;,\\
\jump{\ub^\veps} & = 0\;,\quad \mbox{ on } \Gamma_*^\veps\;,\\
\vphi^\veps & = \bar \vphi^k\;,\quad \mbox{ on } \Gamma_*^{k,\veps}\;,\\
\int_{\Gamma_*^k} \nb \cdot \vec D^\veps \dS & = 0\;,\quad k = 1,2,\dots k^*\;,
\end{split}
\end{equation}
where $\Gamma_{mc}^\veps = \ol{\Om_m^\veps} \cap \ol{\Om_c^\veps}$, and $\nb$ is
the unit normal vector on the interface;   $\nb^\mx$ points outward to
$\Om_m^\veps$.

\item Boundary conditions:
\begin{equation}\label{eq-4b}
\begin{split}
\nb^\mx \cdot \sigmabf^\veps & = \hb^\veps\;, \quad \mbox{ on } \Gamma_\sigma^\veps\;, \\
\nb^\mx \cdot \vec D^\veps  & = \vrho_E^\veps\;, \quad \mbox{ on } \Gamma_{\vec D}^\veps\;, \\
\ub^\veps & = \bar\ub\;, \quad \mbox{ on } \Gamma_{\ub}^\veps\;, \\
\vphi^\veps & = \bar\vphi^0\;, \quad \mbox{ on } \Gamma_{\vphi}^\veps\;, \\
\end{split}
\end{equation}
where $\hb^\veps$ and $\vrho_E^\veps$ are the
applied surface-forces and the surface electric charge, respectively.
\end{itemize}

\begin{remark}\label{rem1}
Two  BVPs will be pursued in what follows:
\begin{enumerate}
\item Case CFD*: the BVP is constituted by \eq{eq-1},\eq{eq-2},\eq{eq-3b},
\eq{eq-4a} and \eq{eq-4b} where the constants $\bar \vphi^k$ for
$k=1,2,\dots,\bar k^\veps$ become  parts of the solution.
\item Case DFC*: the BVP is constituted by \eq{eq-1},\eq{eq-2},\eq{eq-3a},
\eq{eq-4a} and \eq{eq-4b} whereby we shall assume:
\begin{itemize}
\item given values $\bar \vphi^k$ for the finite number on the conductors,
$k=1,\dots,k^*$;
\item vanishing boundary $\Gamma_{\vphi}^\veps$, \ie no prescribed voltage
$\bar\vphi^0$ on the external boundary.
\end{itemize}

\end{enumerate}
The other combinations, namely CFC* and DFD* can be deduced and are left as an
exercise for interested readers.

To obtain a~priori estimates independently of $\veps$, the surface charge
distributed on the interfaces $\Gamma_c^\veps$ must be scaled appropriately by
$\veps$; we shall consider $\varrho_E^\veps = \veps\bar\rho_E$ on
$\Gamma_c^\veps$, where $o(\bar\rho_E) = \veps^0$, see \eq{eq-CFD*0} below.
\end{remark}

\paragraph{Weak formulation}
Since the homogenization procedure is based on the weak formulations, we shall
need the following admissibility sets:
\begin{equation}\label{eq-5a}
\begin{split}
\Ucalbf(\Om_{m*}^\veps) & = \{\vb \in \Hdb(\Om_{m*}^\veps)|\; \vb = \bar\ub \mbox{ on }  \Gamma_{\ub}^\veps\}\;,\\
\Vcal_0(\Om_m^\veps,\Gamma_S^\veps) & = \{\vphi \in H^1(\Om_m^\veps)|\,\vphi = 0
\mbox{ on } \Gamma_S^\veps\}\;,\\
\Vcal_*(\Om_m^\veps,\Gamma_*^\veps) & = \{\psi \in H^1(\Om_m^\veps)|\,\psi = \bar \psi^k
\mbox{ on } \Gamma_*^{k,\veps}\,\;,  k = 1,2,\dots, k^*\}\;\\
\end{split}
\end{equation}
where $\bar \psi^k$ is an arbitrary constant for each $k=1,2,\dots, k^*$. Below
we shall consider two situations which we discuss in Remark~\ref{rem0}; if
$\bar \psi := \bar \vphi$ is prescribed in the definition of
$\Vcal_*(\Om_m^\veps,\Gamma_*^\veps)$,  we write
$\Vcal_*(\Om_m^\veps,\Gamma_*^\veps,\bar \vphi)$.

By $\Ucalbf_0(\Om_{m*}^\veps)$ we denote the space of virtual displacements
derived from $\Ucalbf(\Om_{m*}^\veps)$ for $\bar\ub = \bmi{0}$. Further, we
shall employ the following sets
\begin{equation}\label{eq-5c}
\begin{split}
\Wcal_0(\Om_m^\veps) & = \Vcal_*(\Om_m^\veps,\Gamma_*^\veps)\cap \Vcal_0(\Om_m^\veps,\Gamma_\vphi^\veps)\;,\\
\Wcal_*(\Om_m^\veps,\Gamma_\vphi^\veps,\bar\vphi^0) & = \Vcal_*(\Om_m^\veps,\Gamma_*^\veps) + \hat\vphi\;,\\
\mbox{ where } \hat\vphi \in H^1(\Om_m^\veps),  \mbox{ and } \hat\vphi & = \left\{
\begin{array}{ll}
\bar\vphi^0 &  \mbox{ on } \Gamma_{\vphi}^\veps\;,\\
 0 & \mbox{ on } \Gamma_*\;.
\end{array}
\right.
\end{split}
\end{equation}

Now the weak formulation for Case CFD* can be established: given volume
force $\fb^\veps$ and charge $q_E^\veps$  and the  functions involved in the
boundary conditions \eq{eq-4b}, find $(\ub^\veps,\vphi^\veps,\bar p^\veps) \in
\Ucalbf(\Om_{m*}^\veps) \times
\Wcal_*(\Om_m^\veps,\Gamma_\vphi^\veps,\bar\vphi^0)\times \RR$ such that:
\begin{equation}\label{eq-5}
\begin{split}
\int_{\Om_m^\veps} [\Aop^\veps \eeb{\ub^\veps} - (\parg^\veps)^T\cdot\nabla \vphi^\veps]: \eeb{\vb} \, \dV & + \int_{\Om_*^\veps} [\Aop^\veps \eeb{\ub^\veps}]: \eeb{\vb} \dV \\
- \bar p^\veps \int_{\Gamma_c^\veps} \nb^\cx \cdot \vb\,\dS & =
\int_{\Gamma_\sigma^\veps}\hb^\veps \cdot \vb\dS  + \int_{\Om_{m*}^\veps} \fb^\veps \cdot \vb  \, \dV \;, \\
\int_{\Om_m^\veps} [\parg^\veps:\eeb{\ub^\veps} + \db^\veps \cdot\nabla \vphi^\veps]\cdot\nabla\psi \, \dV& =  \int_{\Om_m^\veps}q_E^\veps\psi \dV + \int_{\Gamma_{\vec D}^\veps\cup \Gamma_c^\veps}\vrho_E^\veps \psi \dS \;,\\
\int_{\pd \Om_c^\veps} \wtilde{\ub^\veps}\cdot \nb^\cx \dS + \gamma^\alpha \bar p^\veps |\Om_c^\veps| & = -J^\veps\;,
\end{split}
\end{equation}
for all $(\vb,\psi) \in \Ucalbf_0(\Om_{m*}^\veps)\times
\Wcal_0(\Om_m^\veps,\Gamma_\vphi^\veps)$. Note that by virtue of the set
$\Vcal_*(\Om_m^\veps,\Gamma_*^\veps)$ defined in \eq{eq-5a}, the solution
$\vphi^\veps$ satisfies the constraint \eq{eq-4a}$_{5,6}$. We recall that the
potential $\bar\vphi^0$ is prescribed on a nonvanishing part of
$\Gamma_{\vphi}^\veps$, see \eq{eq-5a}, which is necessary to ensure a unique
solution of \eq{eq-5}. It is worth to note that, in \eq{eq-5}$_1$, the normal
$\nb^\cx$ is employed instead of normals outward \wrt the solid skeleton.

\begin{remark}\label{rem0}
In the above formulation, $\bar \vphi^{k,\veps}$ involved in the set
$\Vcal_*(\Om_m^\veps,\Gamma_*^\veps)$ are considered as unknown potentials
associated with the conductor parts, whereby, in the sense of test functions,
the corresponding test potential, say  $\bar \psi^{k,\veps}$ can be associated
arbitrary values. However, as a modification of the problem, $\bar
\vphi^{k,\veps}$ could also be considered as given constants. Then the only
difference in the formulation is that $\vphi^\veps \in
\Vcal_*(\Om_m^\veps,\Gamma_*^\veps,\bar\vphi)$ and correspondingly  the set of
the test potentials is replaced by $\Vcal_0(\Om_m^\veps,\Gamma_*^\veps)$; recall
that also   $\Gamma_{\vphi}^\veps = \emptyset$ in this case.
\end{remark}

In this paper we shall focus on the two combination of the cases listed above,
namely CFD* and DFC*. For both these types of microstructures we first restrict
to the no flow situations, thus, considering uniform pressure in each connected
domain filled with fluid.

\section{Homogenization of the static problem -- case CFD*}\label{sec-CFD*}

In this section we are concerned with a steady state of the porous medium which
is in the thermodynamic equilibrium characterized by static fluid in the channels and vanishing
fluid pressure gradients.

The homogenization methods based on the two scale convergence or the
unfolding operator techniques \cite{Cioranescu2008a} can be
applied to describe the limit models arising from asymptotic analyses
of the problem \eq{eq-5} for $\veps\rightarrow 0$.
Although, in this paper, we skip presentation of some of the mathematical
analysis which has been done to prove the convergence result, we explain the
main steps in the derivation of the local problems for computing the so-called
characteristic responses and the macroscopic model equations.

\subsection{Representative periodic cell and its decomposition}

In accordance with the decomposition \eq{eq-pzm1} of domain $\Om$, the fluid
saturated porous medium is   generated by a reference periodic cell $Y$
decomposed into three non-overlapping subdomains $Y_m$, $Y_c$ and $Y_*$, see
Fig.~\ref{fig-CFDF},
\begin{equation}\label{eq-6}
\begin{split}
Y & = Y_m \cup Y_c \cup Y_* \cup
\Gamma_Y\;,\\
 Y_i \cap Y_j & = \emptyset \mbox{ for } j\not = i \mbox{ with } i,j \in \{c,m,*\}\;,\\
Y_* & =\bigcup_k Y_*^k\;,\quad Y_*^k\cap Y_*^l = \emptyset \mbox{ for } k\not = l\;,\\
\dist{Y_*^k}{Y_*^l} & \geq s^*(1-\delta_{kl})\;,\quad s^*>0\;,
\end{split}
\end{equation}
where $s^*$ is the minimum distance of the conductive parts and $\Gamma_Y$,
representing the union of all the interfaces, splits in three disjoint parts (in
the sense of the surface measure),
\begin{equation}\label{eq-6a}
\begin{split}
\Gamma_Y & = \Gamma_{mc} \cup \Gamma_{m*} \cup \Gamma_{c*}\;,\\
\Gamma_{mc} & = \ol{Y_m} \cap \ol{Y_c}\;,\\
 \Gamma_{m*}^k & = \ol{Y_*^k} \cap \ol{Y_m}\quad \mbox{ and } \Gamma_{d*} = \ol{Y_d} \cap \ol{Y_*} \;, \quad d \in \{c,m\}\;.
\end{split}
\end{equation}
For the Case~D* we assume well separated conductor parts, \ie $\ol{Y_*^k}\cap
\pd Y = \emptyset$, such that the lattice generated by one (rescaled) conductor
$Z_*^{k,\veps} = \veps Y_*^k$ forms mutually disconnected inclusions with the
perimeter $\approx \veps$.

We shall need some further notation. By $\Hpdb(Y_m)$ we refer to  the Sobolev
space of vector-valued Y-periodic functions (indicated by the subscript $\#$).
By $\intYs_{D} = |Y|^{-1}\int_D$ with $D\subset \ol{Y}$ we denote the local
average, although $|Y|=1$ can always be chosen. Further we need the space
$\wtilde{\Hpdb}(Y_m)$ which is a restriction of $\Hpdb(Y_m)$ to functions with
the vanishing average, thus, any $\wb \in \wtilde{\Hpdb}(Y_m)$ satisfies
$\intYs_{D}\wb=0$; the space $\wtilde{H_\#^1}(Y_m)$ is defined in analogy. We
also employ $\Pibf^{ij} = (\Pi_k^{ij})$, $i,j,k = 1,2,3$ with components
$\Pi_k^{ij} = y_j\delta_{ik}$.

The following space and sets will be employed in Sections~\ref{sec-CFD*}
and~\ref{sec-DFC*}:
\begin{equation}\label{eq-S*7}
\begin{split}
H_{\#0*}^1(Y_m) &= \{\psi \in H_\#^1(Y_m)|\; \psi = 0  \mbox{ on } \Gamma_{m*}\}\;,\\
H_{\#0,k}^1(Y_m) &= \{\psi \in H_\#^1(Y_m)|\; \psi = \delta_{ki} \mbox{ on } \Gamma_{m*}^i\;,\;i = 1,2,\dots,k^*\}\;,\\
W_{\#*}(Y_m) &= \{\psi\in \wtilde{H_\#^1}(Y_m)|\;\exists (a_k): \psi = \sum_k a_k \hat\vphi^k,\;
\forall \hat\vphi^k \in H_{\#0,k}^1(Y_m)\}.
\end{split}
\end{equation}

\subsection{Asymptotic expansions}

To use the unfolding method of homogenization, or the two-scale convergence
method, a~priori estimates on the solution to the problem \eq{eq-5} should be
obtained. For this, it is necessary to scale appropriately the surface charge
prescribed on the interfaces $\Gamma_{cm}^\veps$, see \eq{eq-4a}$_2$. We assume
constant interface charges $\bar\rho_E$, such that ($\chi_{mc}(y)$ is the
characteristic function of $\Gamma_{mc}$)
\begin{equation}\label{eq-CFD*0}
\Tuf{\vrho_E^\veps(x)} = \veps \bar\rho_E \chi_{mc}(y)\;.
\end{equation}
The a~priori estimates can be derived using the Korn, Poincar\'e and Young
inequalities upon substituting suitable test functions into \eq{eq-5}, so that
obvious manipulations yield the following uniform estimates:
\begin{equation}\label{eq-CFD*1}
\nrm{\ub^\veps}{\Hdb(\Om_m^\veps\cup\Om_*^\veps )} + \nrm{\vphi^\veps}{L^2(\Om_m^\veps)} +
\nrm{\nabla\vphi^\veps}{L^2(\Om_m^\veps)}\leq C\;,
\end{equation}
where the \rhs constant $C$, being independent of $\veps$, depends on all the
data inherited from the boundary conditions, volume forces and electric charges.
As the consequence, the following weak convergences hold:
\begin{equation}\label{eq-CFD*2}
\begin{split}
\Tuf{\ub^\veps} & \cwto \ub^0 \quad \mbox{ in } \Lb^2(\Om\times Y_{m*})\;,\\
\Tuf{\nabla \ub^\veps} & \cwto \nabla_x\ub^0 + \nabla_y \ub^1\quad \mbox{ in } \Lb^2(\Om\times Y_{m*})\;,\\
\Tuf{\vphi^\veps}& \cwto \vphi^0 \quad \mbox{ in } L^2(\Om\times Y_m)\;,\\
\Tuf{\nabla \vphi^\veps}& \cwto \nabla_x\vphi^0 + \nabla_y \vphi^1 \quad \mbox{ in } L^2(\Om\times Y_m)\;,
\end{split}
\end{equation}
where $\ub^1 \in L^2(\Om;\Hpdb(Y_{m*}))$, $\vphi^1\in L^2(\Om;H_\#^1(Y_m))$ and
$\ub^0 \in \Hdb(\Om)$, $\vphi^0 \in H^1(\Om)$. We recall that $\ub^{1\veps}$ and
$\vphi^{1\veps}$ are $Y$-periodic in the second argument.

Due to the interface condition \eq{eq-4a}, it can be proved that
\begin{equation}\label{eq-CFD*3}
\vphi^1(x,\cdot) = \bar\vphi^{k}(x)\quad \mbox{ on } \Gamma_*^k \mbox{ for any } x \in \Om\;,
\end{equation}
therefore $\vphi^1\in L(\Om;W_{\#*}(Y_m))$, see \eq{eq-S*7}. Note that
$\bar\vphi^{k}$ is an unknown field which can be determined once
$\vphi^1(x,\cdot)$ is computed.

By virtue of the asymptotic expansion method, the solution of problem \eq{eq-5}
can be established in the form of the following truncated expansions expressed
in the unfolded forms
\begin{equation}\label{eq-p3}
\begin{split}
\Tuf{\ub^{\veps}} \approx \ub^{R\veps}(x,y) & := \ub^{0\veps}(x) + \veps \ub^{1\veps}(x,y)\;,  \\
\Tuf{\vphi^\veps}\approx\vphi^{R\veps}(x) & := \vphi^{0\veps}(x) + \veps \vphi^{1\veps}(x,y)\;,
\end{split}
\end{equation}
where the local coordinates $y$ are related to the global ones by the mapping
$y_k = \Ycal_k(x)$ defined in Section~\ref{sec-period}. As a consequence,  the
test functions $\vb^\veps$ and $\psi^\veps$ associated with $\ub^{\veps}$ and
$\vphi^\veps$ are considered in the same form of the truncated expansions
\eq{eq-p3} constituted by $\vb^0,\vb^1$, $\psi^0$, and $\psi^1$.

\subsection{Local problems}

Due to the convergence result, the limit in the weak formulation \eq{eq-5} with
test functions $\Tuf{\ub^{\veps}(x)} = \veps \vb^1(x,y)$ and
$\Tuf{\psi^{\veps}(x)} = \veps \psi^1(x,y)$ yield the local equations,
\begin{equation}\label{eq-L1}
\begin{split}
\intY_{\Om \times Y_*}& \nabla_y^S{\vb^1} : \Aop \GrxyS{\ub^0}{\ub^1} \dVxy \\
& + \intY_{\Om \times Y_m}
\nabla_y^S{\vb^1} : [\Aop \GrxyS{\ub^0}{\ub^1} - \parg^T\cdot\Grxy{\vphi^0}{\vphi^1}] \dVxy\\
& = \bar p \int_\Om \intY_{\Gamma_c} \vb^1 \cdot \nb^\cx \dSy \dVx  \;,\\
\intY_{\Om \times Y_m}&\nabla_y\psi^1 \cdot [\parg:\GrxyS{\ub^0}{\ub^1} + \db(\nabla_x \vphi^0 + \nabla_y\vphi^1)]  \dVxy = 0\;,
\end{split}
\end{equation}
for all $\vb^1 \in \Lb^2(\Om;\Hpdb(Y_m))$ and $\psi^1\in L^2(\Om;W_{\#*}(Y_m))$.
\revE{Note that by symbols $\dV$, $\dVxy$, $\dVx$, $\dY$, $\dS$, $\dSx$ and $\dSy$ we
refere to the elementary volumes and surfaces \wrt the different spatial scales.}{}

Due to the linearity of this problem, the fluctuations $\ub^1$ and $\vphi^1$ can
be expressed in terms of the macrsocopic variables:
\begin{equation}\label{eq-L2}
\begin{split}
\ub^1(x,y) & = \omegabf^{ij}e_{ij}^x(\ub^0) + \omegabf^k \pd_k^x \vphi^0 - \bar p \omegabf^P\;, \\ 
\vphi^1(x,y) & = \eta^{ij}e_{ij}^x(\ub^0)+ \eta^k \pd_k^x \vphi^0 - \bar p \eta^P\;,   
\end{split}
\end{equation}
where $\omegabf\in \Hpdb(Y_m)$ and $\eta \in H_\#^1(Y_m)$ are characteristic
responses of displacements and electric potential in the matrix part $Y_m$.

If the structure is perfectly periodic, the decomposition of the microstructure
and the microstructure parameters are independent of the macroscopic position $x
\in \Om$. Otherwise the local problems must be considered at any macroscopic
position, i.e. for almost any $x \in \Om$, cf. \cite{Brown2011}.

We shall use the following bilinear forms:
\begin{equation}\label{eq-L3}
\begin{split}
\aYms{\ub}{\vb} & = \intY_{Y_m \cup Y_*} [\Aop \eeby{\ub}]:\eeby{\vb}\dY\;,\\
\gYm{\ub}{\psi} & = \intY_{Y_m} g_{kij} e_{ij}^y(\ub) \pd_k^y \psi\dY\;,\\
\dYm{\vphi}{\psi} & = \intY_{Y_m} [\db\nabla_y \vphi]\cdot\nabla_y \psi\dY\;.
\end{split}
\end{equation}


The local microstructural response is obtained by solving the following
decoupled problems:
\begin{itemize}
\item Find
$(\omegabf^{ij},\eta^{ij})\in \wtilde{\Hpdb}(Y_m)\times W_{\#*}(Y_m)$ for any $i,j = 1,2,3$
satisfying
\begin{equation}\label{eq-ch1}
\begin{split}
\aYms{\omegabf^{ij} + \Pibf^{ij}}{\vb} - \gYm{\vb}{\eta^{ij}}& = 0\;, \quad \forall \vb \in  \Hpdb(Y_m)\;,\\
\gYm{\omegabf^{ij} + \Pibf^{ij}}{\psi} + \dYm{\eta^{ij}}{\psi}& = 0\;, \quad \forall \psi \in  W_{\#*}(Y_m)\;,
\end{split}
\end{equation}
\item Find
$(\omegabf^{k},\eta^{k})\in \wtilde{\Hpdb}(Y_m)\times W_{\#*}(Y_m)$ for any $k = 1,2,3$
satisfying
\begin{equation}\label{eq-ch2}
\begin{split}
\aYms{\omegabf^{k}}{\vb} - \gYm{\vb}{\eta^{k} + y_k}& = 0\;, \quad \forall \vb \in  \Hpdb(Y_m)\;,\\
\gYm{\omegabf^{k}}{\psi} + \dYm{\eta^{k} + y_k}{\psi}& = 0\;, \quad \forall \psi \in  W_{\#*}(Y_m)\;,
\end{split}
\end{equation}
\item Find
$(\omegabf^P,\eta^P)\in \wtilde{\Hpdb}(Y_m)\times W_{\#*}(Y_m)$
satisfying
\begin{equation}\label{eq-ch3}
\begin{split}
\aYms{\omegabf^P}{\vb} - \gYm{\vb}{\eta^P}& = -\intY_{\Gamma_c} \vb\cdot \nb^\cx \dSy\;, \quad
\forall \vb \in  \Hpdb(Y_m) \;,\\
\gYm{\omegabf^P}{\psi} + \dYm{\eta^P}{\psi}& = 0\;, \quad \forall \psi \in  W_{\#*}(Y_m)\;,
\end{split}
\end{equation}
\end{itemize}

\subsection{Macroscopic model}\label{sec-Mac1}

Using the local corrector basis functions we are able to introduce the
homogenized coefficients which describe the effective poroelastic properties at
the mesoscopic scale. Since the porosity is open
on $\pd \Om$ we need to assume convergence of the boundary segments
$\Gamma_\sigma^\veps,\Gamma_{\vec D}^\veps,\Gamma_{\ub}^\veps$ and
$\Gamma_{\vphi}^\veps$ to $\pd_\sigma \Om$, $\pd_{\vec D} \Om$, $\pd_{u} \Om$ and
$\pd_{\vphi} \Om$. In the limit, the boundary conditions \eq{eq-4b} respected by the
sets \eq{eq-5a} induce the following admissibility sets: $\Ucalbf(\Om) = \{\vb
\in \Hdb(\Om)|\; \vb = \bar\ub \mbox{ on }  \pd_\ub \Om\}$ and
$\Vcal(\Om)=\{\psi \in H^1(\Om)|\psi = \bar\vphi^0 \mbox{ on } \pd_\vphi \Om\}$.
The associated spaces of test functions are $\Ucalbf_0(\Om)$ and $\Vcal_0(\Om)$.
The test  displacements belong to space $\Ucalbf_0(\Om)$ which is restriction of $\Ucalbf(\Om)$ to functions vanishing on $\pd_\ub \Om$. In analogy, the test potentials belong to space $\Vcal_0(\Om)$ which is defined by virtue of $\Vcal(\Om)$ by all functions from $H^1(\Om)$ vanishing on $\pd_\vphi\Om$.

The limit global problem is obtained from \eq{eq-5} with test functions
$\Tuf{\ub^{\veps}(x)} = \veps \vb^0(x)$ and $\Tuf{\psi^{\veps}(x)} = \veps
\psi^0(x)$. The couple $(\ub^0,\vphi^0) \in \Ucalbf(\Om)\times H^1(\Om)/\RR$ and
$\bar p \in \RR$ satisfies
\begin{equation}\label{eq-H1}
\begin{split}
 \int_{\Om} &
e_{ij}^x(\vb^0)\left[\aYms{\ub^1-\Pibf^{kl}e_{kl}^x(\ub^0)}{\Pibf^{ij}} - \gYm{\Pibf^{ij}}{\vphi^1 + y_k\pd_k^x\phi^0}  \right]
 \dVx \\
& - \bar p \int_{\Om} \phi \nabla_x\cdot \vb^0 \, \dVx  =
\int_{{\Om}} \hat\fb \cdot \vb^0 \, \dVx + \int_{\pd {\Om}}\ol{\hb}(\bar p) \cdot \vb^0\dSx\;, \\
\int_{\Om} & \pd_i^x \psi^0 \left[\gYm{\ub^1-\Pibf^{kl}e_{kl}^x(\ub^0)}{y_i} +
\dYm{\vphi^1 + y_k\pd_k^x\phi^0}{y_i}
\right] \dVx \\
& =
\int_{\Om}\hat q_E\psi^0 \dVx + \int_{\pd\Om}\ol{\vrho_E} \psi^0 \dSx \;,\\
\int_{\Om} & \left (\phi\nabla_x\cdot \ub
- \intY_{\Gamma_Y} \ub^{1} \cdot \nb^\mx \dSy \right ) \, \dVx + \bar p \gamma \phi |{\Om}|  = -J\;,
\end{split}
\end{equation}
for all $(\vb^0,\psi^0) \in \Ucalbf_0(\Om)\times H^1(\Om)$, where
$(\ub^1,\vphi^1)$ depends on $(\ub^0,\vphi^0)$ by virtue of the local problem
\eq{eq-L1}. Above the volume charge is $\hat \rho_E := \tilde \rho_E +
|\Gamma_{cm}|/|Y| \bar\rho_E$, where $\tilde \rho_E  =
\intYs_{Y_m}\Tuf{\rho_E^\veps}$. Thus, the constant surface charges defined in
\eq{eq-CFD*0} constitute the effective volume charges involved in the
macroscopic model. Further, $\ol{\vrho_E} = \bar\phi_m \rho_E$ is the effective
surface charge.

The volume forces $\hat\fb$ and boundary tractions $\ol{\hb}(\bar p)$ are
derived in analogy with the treatment explained in
\cite{rohan-etal-CMAT2015-porel}, thus, $\hat\fb = \phi_{m*}\fb$ and
$\ol{\hb}(\bar p) = \bar\phi_{m*}\hb - \nb(1 -\bar\phi_{m*})\bar p$, where
$\phi_{m*}$ and $\bar\phi_{m*}$ are the volume and surface fractions of the solid phase; while $\phi_{m*} = |Y_{m*}|/|Y|$, the surface
fraction is defined {ad hoc}, depending on given assumptions about porosity of
the external surface $\pd\Om$. Obviously, the imposed boundary conditions must
be coherent with these assumptions. It is worth noting that here $\ol{\hb}(\bar
p)$ applies as the limit surface force to the situation of the static loading
with the drained conditions on the external pore boundary $\pd_\ext\Om_c^\veps$.
Obviously, for the case DF, when only fluid inclusions are considered, the
surface porosity is zero, thus, $\bar\phi_{m*} = 1$.

The characteristic responses \eq{eq-ch1}--\eq{eq-ch3} obtained at the
microscopic scale allow us to express the local averaging integrals in
\eq{eq-H1} involving the two-scale functions using the homogenized coefficients.
\revE{Their expressions are identified upon substituting the split \eq{eq-L2} into the three equations in \eq{eq-H1}:
}{}
%
In the equilibrium equation:
\begin{equation}\label{eq-H5a}
\begin{split}
A_{klij}^H & = \aYms{\omegabf^{ij} + \Pibf^{ij}}{\Pibf^{kl}} - \gYm{\Pibf^{kl}}{\eta^{ij}}\\
& = \aYms{\omegabf^{ij} + \Pibf^{ij}}{\omegabf^{kl}+\Pibf^{kl}} + \dYm{\eta^{kl}}{\eta^{ij}}\;, \\
B_{ij}^H & = \aYms{\omegabf^P}{\Pibf^{ij}} - \gYm{\Pibf^{ij}}{\eta^P} + \phi\delta_{ij}\;,\\
G_{kij}^H & = \gYm{\Pibf^{ij}}{\eta^k + y_k} - \aYms{\omegabf^k}{\Pibf^{ij}}\;.
\end{split}
\end{equation}
In the electricity equation:
\begin{equation}\label{eq-H5b}
\begin{split}
\acute G_{kij}^H & = \gYm{\omegabf^{kl} + \Pibf^{kl}}{y_k} + \dYm{\eta^{ij}}{y_k}\;,\\
D_{kl}^H & = \gYm{\omegabf^l}{y_k} + \dYm{\eta^l + y_l}{y_k}\\
& = \dYm{\eta^l + y_l}{\eta^k + y_k} + \aYms{\omegabf^k}{\omegabf^l}\;,\\
\acute F^H_i & = \gYm{\omegabf^P}{y_i} + \dYm{\eta^P}{y_i}\;.
\end{split}
\end{equation}
In the fluid mass conservation equation:
\begin{equation}\label{eq-H5c}
\begin{split}
\acute B_{ij}^H & =  -\intY_{Y_m}\dvg_y \omegabf^{ij}\dY + \phi\delta_{ij}= \aYms{\omegabf^P}{\Pibf^{ij}} - \gYm{\Pibf^{ij}}{\eta^P}+ \phi\delta_{ij}\;,\\
M^H  & = \intY_{\Gamma_Y} \omegabf^P\cdot \nb^\mxe \dSy + \gamma \phi=  \aYms{\omegabf^P}{\omegabf^P} + \dYm{\eta^P}{\eta^P} + \gamma \phi \;,\\
F^H_i & = \intY_{\Gamma_Y}\omegabf^{i}\cdot\nb^\mxe\dSy\;.\\
\end{split}
\end{equation}
It is worth noting that the effective elasticity $\Aop^H = (A_{klij}^H)$
inherits all the symmetry properties of the piezoelectric material elasticity
$\Dop$. Also the other poroelastic coefficients, namely  the symmetric Biot
stress-coupling coefficient $B_{ij}^H$, and the positive Biot compressibility
coefficient $M^H$, reflect the piezoelectric properties of the skeleton. The
expressions proving the symmetry and positivity properties are obtained using
\eq{eq-ch1}-\eq{eq-ch3}.

Now we can rewrite equations  \eq{eq-H1} of the global problem in terms of the
homogenized coefficients \eq{eq-H5a}-\eq{eq-H5c},
\begin{equation}\label{eq-H11b}
\begin{split}
 & \int_{\Om}[\Aop^H\eeb{\ub^0} - (\coefG)^T\nabla \vphi^0 - \bar p \Bb^H
]:\eeb{\vb^0} \dVx =
\int_{{\Om}} \hat\fb \cdot \vb^0 \, \dVx + \int_{\pd {\Om}}\ol{\hb}(\bar p) \cdot \vb^0\dSx\;, \\
& \int_{\Om} [\coefaG\eeb{\ub^0} + \Db^H \nabla \vphi^0 - \coefaF \bar p] \cdot \nabla \psi^0\dVx  =
\int_{\Om}\hat q_E\psi^0 \dVx + \int_{\pd\Om}\ol{\vrho_E} \psi^0 \dSx \;,\\
& \int_{\Om} \left (
\acute\Bb^H:\eeb{\ub^0} - \coefF\cdot \nabla \vphi^0 + M^H  \bar p\right ) \, \dVx  +  \bar p \gamma |\Om| = - J\;.
\end{split}
\end{equation}

In accordance with phenomenological theory, one should expect the Onsager
reciprocity relationships to be satisfied, which is one of the advantageous
features of the homogenization method. Indeed, the following equalities are
{proved in the Appendix~A,}
\begin{equation}\label{eq-H8}
F^H_i = \acute F^H_i\;, \quad
G_{kij}^H = \acute G_{kij}^H\;, \quad
B_{ij}^H = \acute B_{ij}^H \;.
\end{equation}


Using the symmetry relationships \eq{eq-H8}, the macroscopic problem \eq{eq-H11b} can be reformulated, as
follows: Find $(\ub^0,\vphi^0) \in \Ucalbf(\Om)\times\Vcal(\Om)$ and $\bar p \in
\RR$, such that
\begin{equation}\label{eq-H11}
\begin{split}
 & \int_{\Om}[\Aop^H\eeb{\ub^0} - (\coefG)^T\nabla \vphi^0 - \bar p \Bb^H]:\eeb{\vb^0} \dVx =
\int_{{\Om}} \hat\fb \cdot \vb^0 \, \dVx + \int_{\pd_\sigma {\Om}}\ol{\hb}(\bar p) \cdot \vb^0\dSx\;, \\
& \int_{\Om} [\coefG\eeb{\ub^0} + \Db^H \nabla \vphi^0 - \coefF \bar p] \cdot \nabla \psi^0\dVx  =
\int_{\Om}\hat q_E\psi^0 \dVx + \int_{\pd_D\Om}\ol{\vrho_E} \psi^0 \dSx \;,\\
& \int_{\Om} \left (\Bb^H:\eeb{\ub^0} - \coefF\cdot \nabla \vphi^0 + M^H  \bar p\right ) \, \dVx   = - J\;,
\end{split}
\end{equation}
for all $(\vb^0,\psi^0) \in \Ucalbf_0(\Om)\times\Vcal_0(\Om)$.

From \eq{eq-H11}, the strong form of the macroscopic problem can be obtained. We
present a generalized formulation which admits both the cases CF and DF, the
latter deduced for the separated inclusions, thus, giving rise the locally
defined fluid pressure $p(x)$; we are concerned with this case in the next
section. The equilibrium equations and the boundary conditions( we drop the
superscripts $^0$),
\begin{equation}\label{eq-H-eq}
\begin{split}
-\nabla\cdot \sigmabf^H(\ub,\vphi,p) & = \hat\fb \;,\quad \mbox{ in }  \Om\;,\\
 \nabla\cdot \vec D^H(\ub,\vphi,p) & = \hat q_E \;,\quad \mbox{ in }  \Om\;, \\
\sigmabf^H(\ub,\vphi,p)\cdot \nb  & = \ol{\hb}\;,\quad \mbox{ on } \pd_\sigma\Om\;,\\
\quad \vec D^H(\ub,\vphi,p)\cdot \nb  & = -\ol{\vrho_E}\;,\quad \mbox{ on } \pd_D\Om
\end{split}
\end{equation}
involve the effective constitutive equations for the upscaled porous
piezoelectric material:
\begin{equation}\label{eq-H12}
\begin{split}
 \sigmabf^H & = \Aop^H\eeb{\ub} - (\coefG)^T\nabla \vphi- p \Bb^H\;,\\
\vec{D} & = \coefG\eeb{\ub} + \Db^H \nabla \vphi - \coefF  p\;,\\
-p & = \frac{1}{M^H}\left( \Bb^H:\eeb{\ub} - \coefF\cdot \nabla \vphi + j\right)\;,
\end{split}
\end{equation}
where $p = \bar p$ and $j = J/|\Om|$ is the local fluid volume production in the
porous material per volume.

\begin{remark}\label{rem-j}
Although, in this study, we consider static loading such that no pressure
gradients appear, the homogenization result can be extended for quasistatic,
nonstationary problems. To do so, we consider $p$ as a scalar field depending on
$x$, and put $j = \nabla\cdot\wb$, where $\wb$ is the seepage velocity, \cf
\cite{Rohan-AMC}, where an analogous treatment was pursued.
\end{remark}

As the consequence of \eq{eq-H12}, the pressure can be eliminated from the
\eq{eq-H12}$_{1,2}$, so that
\begin{equation}\label{eq-H13}
\begin{split}
 \sigmabf^H & = \Aop^U\eeb{\ub} - (\coefGU)^T\nabla \vphi + (M^H)^{-1}\Bb^H j\;,\\
\vec{D} & = \coefGU\eeb{\ub} + \Db^U \nabla \vphi + (M^H)^{-1}\coefF  j\;,
\end{split}
\end{equation}
where the following coefficients labelled by superscript $^U$ can be considered
as {the ``undrained'' effective material properties:}
\begin{equation}\label{eq-H14}
\begin{split}
\Aop^U & = \Aop^H + (M^H)^{-1}\Bb^H\otimes\Bb^H \quad\mbox{ undrained elasticity},\\
\Db^U & = \Db^H - (M^H)^{-1}\coefF\otimes\coefF \quad\mbox{ undrained dielectricity},\\
\coefGU & = \coefG + (M^H)^{-1}\coefF \Bb^H \quad\mbox{ undrained piezoeletric coupling}.
\end{split}
\end{equation}
It should be noticed that for  $J=0$, \ie in the undrained situation,
\eq{eq-H13} gives the  constitutive law which is analogous to a piezoelectric
solid.
The final remark concerns the electric field generated alternatively by {pore pressure}, or by the { fluid injection}. If the macroscopic strains vanish $\eeb{\ub} = 0$, then 
\begin{equation*}\label{eq-H15a}
\begin{split}
\nabla \vphi = (\Db^H)^{-1} \coefF {\bar p}\;,\quad\mbox{ or }\quad
\nabla \vphi = -(\hat M {{\Db}^U})^{-1} \coefF {j}\;.
\end{split}
\end{equation*}



\section{Homogenization of the static problem -- case DFC*}\label{sec-DFC*}

In this section we treat the problem with prescribed voltage on a finite number
of mutually disconnected conductor networks penetrating into the period
structure of the porous material. As a consequence, if the electric field is
getting stronger with $\veps\rightarrow 0$, also the dielectric properties of
the piezoelectric material must be decreasing in the right order, so that the
electric displacements remain bounded.

\subsection{Microstructures and material scaling}

We consider formulation \eq{eq-5} with given potentials $\bar\vphi^k$ for each
simply connected domain $\Om_*^{k,\veps}$ occupied by the perfect conductor and
represented by $Y_*^k$ within the cell $Y$. We shall assume that on the external
boundary of the piezoelectric structure no  voltage is prescribed, thus,
$\Gamma_\vphi^\veps = \emptyset$, see \eq{eq-pzm1a} and \eq{eq-4b}. The
following two cases can be considered:
\begin{list}{}{} 
\item (W) Weakly controlled field: $\bar\vphi^{k,\veps} = \veps \bar\vphi^{k}$;
\item (S) Strongly controlled field: $\bar\vphi^{k,\veps} = \bar\vphi^{k}$.
\end{list} 
In both these cases, $\bar\vphi^{k}$ is independent of $\veps$. It can be shown
that the convergence result related to the potential $\vphi^\veps$ of the model
treated above in Section~\ref{sec-CFD*} can be adapted easily for the case
DFC*W, which leads to the same limit homogenized mode, as the one introduced
before. Therefore, here we focus on the media with strongly controlled
potentials, namely on the case DFC*S.

Since $\bar\vphi^{k}$ does not vanish with $\veps\rightarrow 0$, steep gradients on
the electric potential are assumed for small $\veps$. As the consequence, to
preserve finite electric field in the limit, we consider the following scaling
of the dielectric and piezoelectric coefficients:
\begin{equation}\label{eq-S*1}
\begin{split}
\left.
\begin{array}{ll}
\parg^\veps(x) & = \veps \bar\parg\;, \\
\db^\veps(x) & = \veps^2 \bar\db\;,
\end{array}\right\}
\quad \mbox{ in } \Om_m^\veps\;.
\end{split}
\end{equation}
Since we consider Case DF, \ie $\Om_c^\veps$ is represented by all
inclusions $\Om_c^{j,\veps}$, in the no-flow condition, the pressure is
distributed as a piecewise constant function attaining a constant value in any
$\Om_c^{j,\veps}$. Therefore, we define the space $\Pcal(\Om_c^\veps) = \{ p \in
L^2(\Om_c^\veps)|\; p = \bar p^j \mbox{ in } \Om_c^{j,\veps}\}$ with $\bar p^j
\in \RR$ is representing any constant. Obviously, the number of inclusions
increases with $\veps^{-3}$. As the consequence of the DF type porosity, we
assume no pore intersects boundary $\pd \Om$ so that  neither $\Gamma_\sigma$,
nor $\hb^\veps$ depend on $\veps$ in the boundary condition specified in
\eq{eq-4b}.

Given the potential values $\bar\vphi^0=\{\bar \vphi^{k}\}$, $k= 1,2,\dots,k^*$
in each subdomain $\Om_*^{k,\veps}$ of $\Om_*^\veps$, volume force $\fb^\veps$,
volume charge $q_E^\veps$,  and the  functions involved in the  \rhs of \eq{eq-4b},
find $(\ub^\veps,\vphi^\veps,p^\veps) \in \Ucalbf(\Om_m^\veps) \times
\Vcal_*(\Om_m^\veps,\Gamma_*^\veps,\bar\vphi)\times \Pcal(\Om_c^\veps)$ such
that:
\begin{equation}\label{eq-5*}
\begin{split}
\int_{\Om_m^\veps} & [\Aop^\veps \eeb{\ub^\veps} - \veps(\bar\parg)^T\cdot\nabla \vphi^\veps]: \eeb{\vb} \, \dV  + \int_{\Om_*^\veps} [\Aop^\veps \eeb{\ub^\veps}]: \eeb{\vb}\dV  \\
& -  \int_{\Gamma_c^\veps} p^\veps\nb^\cx \cdot \vb\,\dS  =
\int_{\Gamma_\sigma}\hb \cdot \vb\dS  + \int_{\Om_{m*}^\veps} \fb^\veps \cdot \vb  \, \dV \;, \\
\int_{\Om_m^\veps} & [\veps\bar\parg:\eeb{\ub^\veps} + \veps^2\bar\db \cdot\nabla \vphi^\veps]\cdot\nabla\psi \, \dV =  \int_{\Om_m^\veps}q_E^\veps\psi \dV + \int_{\Gamma_{\vec D}^\veps\cup \Gamma_c^\veps}\vrho_E^\veps \psi \dS \;,\\
\int_{\pd \Om_c^\veps}& q \wtilde{\ub^\veps}\cdot \nb^\cx \dS + \gamma^\alpha \int_{\Om_c^\veps} p^\veps q  \, \dV = 0\;,
\end{split}
\end{equation}
for all $(\vb,\psi,q) \in \Ucalbf_0(\Om_m^\veps)\times
\Vcal_0(\Om_m^\veps,\Gamma_*^\veps) \times \Pcal(\Om_c^\veps)$.

\subsection{Asymptotic expansions}

We proceed in analogy with derivation of the CFD* model reported in Section
\ref{sec-CFD*}. First we obtain the  a~priori estimates. Recalling the scaling
ansatz  \eq{eq-CFD*0} for the surface charge and the scaling \eq{eq-S*1}
concerning piezoelectric  material coefficients, standard manipulations in
\eq{eq-5*} yield
\begin{equation}\label{eq-S*2}
\begin{split}
\nrm{\ub^\veps}{\Hdb(\Om_m^\veps\cup \Om_*^\veps)} + \nrm{\vphi^\veps}{L^2(\Om_m^\veps)} +
\veps\nrm{\nabla\vphi^\veps}{L^2(\Om_m^\veps)}
+ \nrm{p^\veps}{L^2(\Om_c^\veps)} \leq C\;,
\end{split}
\end{equation}
where the constant $C$, being independent of $\veps$, reflects the data of the
problem.

To respect the boundary conditions, recalling $\bar\vphi^{k}$ are given
constants for $k = 1,\dots,\bar k^\ast$, we require
\begin{equation}\label{eq-S*4}
\vphi^\veps = \bar\vphi^{k} \quad \mbox{ on } \Om\times \Gamma_{m*}^k\;,
\end{equation}
where $\Gamma_{m*}^k$ is defined in \eq{eq-6a}.

Due to \eq{eq-S*2}, the following convergences hold
\begin{equation}\label{eq-S*5}
\begin{split}
\Tuf{\ub^\veps} & \cwto \ub^0 \quad \mbox{ in } \Lb^2(\Om\times Y_{m*})\;,\\
\Tuf{\nabla \ub^\veps} & \cwto \nabla_x\ub^0 + \nabla_y \ub^1\quad \mbox{ in } \Lb^2(\Om\times Y_{m*})\;,\\
\Tuf{\vphi^\veps}& \cwto \hat\vphi^0 \quad \mbox{ in } L^2(\Om\times Y_m)\;,\\
\veps\Tuf{\nabla \vphi^\veps}& \cwto \nabla_y\hat\vphi^0 \quad \mbox{ in } L^2(\Om\times Y_m)\;,\\
\Tuf{p^\veps} & \cwto p^0\quad \mbox{ in } L^2(\Om\times Y_c)\;.\\
\end{split}
\end{equation}
Moreover, due to the trace theorem, the electric two-scale potential satisfies
the conditions
\begin{equation}\label{eq-S*6}
\hat\vphi^0(x,\cdot) = \bar\vphi^{k}(x)\quad \mbox{ on } \Gamma_*^k \mbox{ for any } x \in \Om\;,
\end{equation}
where $\bar\vphi^{k}$ are given, $k = 1,2,\dots,k^*$. This makes the difference
with treatment in Section~\ref{sec-CFD*}, see \eq{eq-CFD*3}, dealing with the
case CFD*, there  by $\bar\vphi^{k}(x)$ we mean an unknown value of the
potential attained on the interface $\Gamma_*^{k}$, thus, $\bar\vphi^{k}$ is a
part of the homogenized problem solution, contrary to the present situation.

We shall employ the space $H_{\#0*}^1(Y_m)$ and the set $H_{\#0,k}^1(Y_m)$
introduced in \eq{eq-S*7}. Now \eq{eq-S*5} and \eq{eq-S*6} yield
$\hat\vphi^0(x,\cdot) \in V_{\bar\vphi}(Y_m) := H_{\#0*}^1(Y_m) + \sum_k
\chi_*^k \hat\vphi^k(x,\cdot) \bar\vphi^k(x)$ for any $x\in \Om$ with
$\hat\vphi^k(x,\cdot) \in H_{\#0,k}^1(Y_m)$.

The convergence result enables to introduce formally the following truncated
expansions of displacements and potential such that the limit equations can be
derived:
\begin{equation}\label{eq-S*3}
\begin{split}
\Tuf{\ub^{\veps}} \approx \ub^{R\veps}(x,y) & := \ub^{0\veps}(x) + \veps \ub^{1\veps}(x,y)\;,  \\
\Tuf{\vphi^\veps}\approx\vphi^{R\veps}(x,y) & := \hat\vphi^{0\veps}(x,y) \;,\\
\Tuf{p^\veps}\approx p^{R\veps}(x,y) & := p^{0\veps}(x)\;.
\end{split}
\end{equation}
where $\ub^{1\veps}$ and $\Phi^{0\veps}$ are $Y$-periodic in the second
argument. By virtue of \eq{eq-S*5}, we assume the convergence of
$\ub^{0,\veps},\ub^{1\veps}$,  $\Phi^{0\veps}$ and $p^{0\veps}$ to the limit
functions $\ub^{0},\ub^{1}$,  $\hat\vphi^{0}$ and $p^{0}$, respectively. In
analogy with \eq{eq-S*3}, the test functions $\vb^\veps$, $\hat\psi^\veps$ and
$q$ are expressed in terms of $\vb^0,\vb^1$, $\hat\psi^0$ and $q^0$ which are
associated with $\ub^{0\veps},\ub^{1,\veps}$, $\vphi^{0\veps}$ and $p^\veps$,
respectively.

\subsection{Local problems}

We recall the notation of the fluid-solid interface: $\Gamma_c = \Gamma_{mc}
\cup \Gamma_{c*}$ with the fluid-outward normal $\nb^\cx$. Straightforward
calculations lead to
\begin{equation}\label{eq-L*S1}
\begin{split}
\intY_{\Om \times (Y_m \cup Y_*)}&
\eeby{\vb^1} : \Aop \GrxyS{\ub^0}{\ub^1}\dVxy
- \intY_{\Om \times Y_m }\eeby{\vb^1} :\bar\parg^T\nabla_y\hat\vphi^0 \dVxy\\
& =  \int_\Om p^0 \intY_{\Gamma_c} \vb^1 \cdot \nb^\cx \dSy \dVx  \;,\\
\intY_{\Om \times Y_m}&\nabla_y\hat\psi \cdot [\bar\parg:\GrxyS{\ub^0}{\ub^1} + \bar\db\nabla_y\hat\vphi^0]\dVxy = \intY_{\Om \times \Gamma_{mc}}\bar\rho_E \hat\psi\dVxy\;,
\end{split}
\end{equation}
which holds for all $\vb^1 \in \Lb^2(\Om;\Hpdb(Y_m))$ and $\hat\psi\in
L^2(\Om;W_{\#*}(Y_m)$.

The two-scale functions can be expressed in terms of the characteristic
responses $(\omegabf,\eta)$, such that
\begin{equation}\label{eq-L2b}
\begin{split}
\ub^1(x,y) & = \omegabf^{ij}e_{ij}^x(\ub^0)  - p^0 \omegabf^P + \omegabf^\rho \rho_E + \sum_k\hat\omegabf^k\bar\vphi^k\;,\\
\vphi^0(x,y) & = \hat\eta^{ij}e_{ij}^x(\ub^0) - p^0 \hat\eta^P + \hat\eta^\rho \rho_E + \sum_k\hat\vphi^k\bar\vphi^k\;.
\end{split}
\end{equation}
All $(\omegabf,\eta)$ are Y-periodic, representing the  displacements in the
entire solid part, $Y_{m*} = Y_m \cup Y_*$ and the electric potential in the
matrix part $Y_m$.

We shall use the  bilinear forms \eq{eq-L3} with obvious modifications, namely
$\gYm{\cdot}{\cdot}$ and $\dYm{\cdot}{\cdot}$ involve the coefficients $\bar
g_{kij}$ and $\bar d_{kl}$, respectively, see \eq{eq-S*1}.

The local microstructural response is obtained by solving the following
decoupled problems:
\begin{itemize}
\item Find
$(\omegabf^{ij},\hat\eta^{ij})\in \Hpdb(Y_{m*})\times H_{\#0*}^1(Y_m)$ for any $i,j = 1,2,3$
satisfying
\begin{equation}\label{eq-ch1S}
\begin{split}
\aYms{\omegabf^{ij} + \Pibf^{ij}}{\vb} - \gYm{\vb}{\hat\eta^{ij}}& = 0\;, \quad \forall \vb \in  \Hpdb(Y_{m*})\;,\\
\gYm{\omegabf^{ij} + \Pibf^{ij}}{\psi} + \dYm{\hat\eta^{ij}}{\psi}& = 0\;, \quad \forall \psi \in  H_{\#0*}^1(Y_m)\;,
\end{split}
\end{equation}
\item Find
$(\omegabf^P,\hat\eta^P)\in \Hpdb(Y_{m*})\times H_{\#0*}^1(Y_m)$
satisfying
\begin{equation}\label{eq-ch3S}
\begin{split}
\aYms{\omegabf^P}{\vb} - \gYm{\vb}{\hat\eta^P}& = -\intY_{\Gamma_{c}} \vb\cdot \nb^\cx \dSy\;, \quad
\forall \vb \in  \Hpdb(Y_m) \;,\\
\gYm{\omegabf^P}{\psi} + \dYm{\hat\eta^P}{\psi}& = 0\;, \quad \forall \psi \in  H_{\#0*}^1(Y_m)\;,
\end{split}
\end{equation}
\item Find
$(\omegabf^\rho,\hat\eta^\rho)\in \Hpdb(Y_{m*})\times H_{\#0*}^1(Y_m)$
satisfying
\begin{equation}\label{eq-ch4S}
\begin{split}
\aYms{\omegabf^\rho}{\vb} - \gYm{\vb}{\hat\eta^\rho}& = 0\;, \quad
\forall \vb \in  \Hpdb(Y_m) \;,\\
\gYm{\omegabf^\rho}{\psi} + \dYm{\hat\eta^\rho}{\psi}& = \intY_{\Gamma_{mc}} \psi \dSy\;, \quad \forall \psi \in  H_{\#0*}^1(Y_m)\;,
\end{split}
\end{equation}
\item Find
$(\hat\omegabf^k,\hat\vphi^k)\in \Hpdb(Y_{m*})\times H_{\#0,k}^1(Y_m)$
satisfying, for $k = 1,2,\dots,k^*$,
\begin{equation}\label{eq-ch5S}
\begin{split}
\aYms{\hat\omegabf^k}{\vb} - \gYm{\vb}{\hat\vphi^k}& = 0 \;, \quad
\forall \vb \in  \Hpdb(Y_m) \;,\\
\gYm{\hat\omegabf^k}{\psi} + \dYm{\hat\vphi^k}{\psi}& = 0\;, \quad \forall \psi \in  H_{\#0*}^1(Y_m)\;.
\end{split}
\end{equation}
\end{itemize}

\subsection{Macroscopic model}\label{sec-Mac2}

We pursue the analogous procedure reported in Section~\ref{sec-Mac1}. As the
consequence of the convergence result related to the two-scale potential
$\hat\vphi^0(x,y)$, the limit problem is defined in terms of $\ub^0$ and $p^0$
only: Find $\ub^0 \in \Ucalbf(\Om)$ and $p^0\in L^2(\Om)$ such that
\begin{equation}\label{eq-H1*S}
\begin{split}
\int_{\Om} &
e_{ij}^x(\vb^0)\left[\aYms{\ub^1-\Pibf^{kl}e_{kl}^x(\ub^0)}{\Pibf^{ij}} - \gYm{\Pibf^{ij}}{\hat\vphi^0 }  \right] \dVx \\
& -\int_{\Om} p^0 \phi \nabla_x\cdot \vb^0 \, \dVx  =
\int_{{\Om}} \hat\fb \cdot \vb^0 \, \dVx + \int_{\pd {\Om}}\hb \cdot \vb^0\dSx\;, \\
\int_{\Om}&  q^0\left(\phi\nabla_x\cdot \ub^0
- \intY_{\Gamma_c} \ub^{1} \cdot \nb^\mxe \dSy \right ) \, \dVx +\gamma \int_{\Om}\phi  p^0 q^0\, \dVx = 0\;,
\end{split}
\end{equation}
for all $(\vb^0,q^0) \in \Ucalbf_0(\Om)\times L^2(\Om)$.
By $\hat\fb = \intYs_{Y_{m*}} \fb$ the average volume forces are denoted.

Using the characteristic responses \eq{eq-ch1S}--\eq{eq-ch5S} obtained at the
microscopic scale, upon substituting the split form of the two-scale functions
\eq{eq-L2b} into \eq{eq-H1*S}, the homogenized coefficients can be computed in
analogy with the treatment explained in Section~\ref{sec-CFD*}.
\begin{equation}\label{eq-H*S}
\begin{split}
A_{klij}^H & = \aYms{\omegabf^{ij} + \Pibf^{ij}}{\Pibf^{kl}} - \gYm{\Pibf^{kl}}{\hat\eta^{ij}}\\
& = \aYms{\omegabf^{ij} + \Pibf^{ij}}{\omegabf^{kl}+\Pibf^{kl}} + \dYm{\hat\eta^{kl}}{\hat\eta^{ij}}\;, \\
B_{ij}^H & = \aYms{\omegabf^P}{\Pibf^{ij}} - \gYm{\Pibf^{ij}}{\hat\eta^P} + \phi\delta_{ij} =
-\intY_{Y_m} \nabla_y\cdot \omegabf^{ij}\dY + \phi\delta_{ij}
\;,\\
H_{ij}^k & = \aYms{\hat\omegabf^k}{\Pibf^{ij}} -  \gYm{\Pibf^{ij}}{\hat\vphi^k}\;,\\
S_{ij}^H & = \aYms{\omegabf^\rho}{\Pibf^{ij}} -  \gYm{\Pibf^{ij}}{\hat\eta^\rho}\;,\\
R^H & = - \intY_{\Gamma_{c}} \omegabf^\rho\cdot \nb^\cx\dSy\;,\\
M^H & = \intY_{\Gamma_c} \omegabf^P\cdot \nb^\cx \dSy + \phi\delta_{ij}
= \aYms{\omegabf^P}{\omegabf^P} + \dYm{\hat\eta^P}{\hat\eta^P}
 + \phi\delta_{ij}\;,\\
Z^k & = - \intY_{\Gamma_c}\hat\omegabf^k\cdot \nb^\cx \dSy\;.
\end{split}
\end{equation}

We can now rewrite \eq{eq-H1*S} in terms of the coefficients \eq{eq-H*S},
thus, the macroscopic problem for the DFC* reads: Find $\ub^0 \in \Ucalbf(\Om)$ and $p^0\in L^2(\Om)$ such that
\begin{equation}\label{eq-H11b*S}
\begin{split}
 \int_{\Om}\eeb{\vb^0}:\left(\Aop^H\eeb{\ub^0} - p \Bb^H\right) \dVx =
& - \int_{\Om}\eeb{\vb^0}:\left(\sum_k \Hb^k \bar\vphi^k + \Sb^H \rho_E \right)\dVx\\
& + \int_{\Om} \hat\fb \cdot \vb^0\dVx  + \int_{\pd {\Om}}\ol{\hb}(\bar p) \cdot \vb^0\dSx
\;,\\
 \int_{\Om} q^0\left(\Bb^H:\eeb{\ub^0} + p M^H\right)  \dVx =
&  \int_{\Om}q^0\left(\sum_k Z^k \bar\vphi^k + R^H \rho_E \right)\dVx\;,
\end{split}
\end{equation}
for all $\vb^0 \in \Ub_0(\Om)$ and for all $q^0 \in L^2(\Om)$.

The homogenized piezoelectric material obeys the poroelastic law with the
poroelastic coefficients modified due to the piezoelectric effect. The imposed
voltage, \ie the  potentials prescribed on the distributed conductors, and the
interface charges generate a prestress which, upon integrating by parts,
can be presented in terms of volume forces.

Analogous procedure pursued for the case DFC*S leads to the same poroelastic
model, however, since the fluid inclusions are disconnected, $\ol{Y_c} \subset
Y$, the macroscopic pressure $p(x)$ depends on $x$. To simplify the notation, in
what follows we drop the superscripts $^0$ in all macroscopic variables. In
analogy with \eq{eq-H-eq} and \eq{eq-H12}, the differential form of
\eq{eq-H11b*S} is presented by the following equations, dropping the
superscripts $^0$:
\begin{equation}\label{eq-H*S1}
\begin{split}
-\nabla\cdot \sigmabf^H(\ub,p) & = \hat\fb \;,\quad \mbox{ in }  \Om\;,\\
\sigmabf^H(\ub,p)\cdot \nb  & = \ol{\hb}\;,\quad \mbox{ on } \pd_\sigma\Om\;,
\end{split}
\end{equation}
with the constitutive equations:
\begin{equation}\label{eq-H*S2}
\begin{split}
  \sigmabf^H(\ub,p) & = \Aop^H\eeb{\ub} - p \Bb^H
+ \sum_k \Hb^k \bar\vphi^k + \Sb^H \rho_E \;,\\
 p & = \frac{1}{M^H}\left( \sum_k Z^k \bar\vphi^k + R^H \rho_E - \Bb^H:\eeb{\ub}- j\right)\;,
\end{split}
\end{equation}
where $j = J/|\Om|$ has been introduced in \eq{eq-H12} assuming the connected
porosity; for disconnected fluid-saturated inclusions we put $j = 0$ in
\eq{eq-H*S2}. From \eq{eq-H*S2}$_1$, the pressure can be eliminated, so that the
``undrained'' coefficients can be obtained in analogy with \eq{eq-H14}.

\begin{remark}\label{rem10}
Both the models, as presented in Sections~\ref{sec-CFD*} and \ref{sec-DFC*} can
be adapted easily for the other type of the porosity, thus, the form of the
macroscopic constitutive  equations can be  deduced easily also for cases CFC*
and DFD*. If the porosity is connected, the models can be extended for fluid
flow problems; for this, the pressure is considered as a scalar field $p(x)$
and,  in \eq{eq-H12} and \eq{eq-H*S2}, we put $j = \nabla\cdot\wb$, where,
denoting by $\vb^f$ the mean fluid velocity, $\wb = \phi(\vb^f - \dot\ub)$ is
the seepage velocity governed by the Darcy flow law  $\wb  =
-\bar\eta^{-1}\Kb\nabla p$ involving the fluid viscosity. As far as the
quasistatic problems are considered, cf. \cite{RSW-ComGeo2013}, the permeability
can be obtained for a specific geometry by the standard homogenization of the
Stokes flow, see \eg \cite{Hornung-Book1996}.
\end{remark}

\section{Problem formulation for the homogenized media}

To allow for more general situations of loading the deforming piezoporoelastic
media, we shall generalize the homogenization results obtained above for the two
different cases of heterogeneous media by considering more general boundary
conditions. To introduce them for the coupled problem, we need  the following
two decompositions of $\pd\Om$ into disjoint parts:
\begin{equation}\label{eq-Om-split}
\begin{split}
\pd \Om & = \pd_\sigma \Om \cup \pd_u \Om\;, \quad \pd_\sigma \Om \cap \pd_u \Om = \emptyset\;, \\
\pd \Om & = \pd_E \Om \cup \pd_\vphi \Om\;, \quad \pd_E \Om \cap \pd_\vphi \Om = \emptyset\;.
\end{split}
\end{equation}
The displacement and  the electric potential 
must satisfy the
boundary conditions which will now be specified for the two homogenized models derived above.

\subsection{Model CFD*} 

We consider the following boundary conditions, where  $D_n$ and $\gb^s$ are given:
\begin{equation}\label{eq-B2}
\begin{split}
\ub = \bmi{0} \;\quad \mbox{ on } \pd_u\Om\;,\quad \quad \nb\cdot\sigmabf = \gb^s\;\quad \mbox{ on } \pd_\sigma\Om\;,\\
\vphi = \bar \vphi^0 \;\quad \mbox{ on } \pd_\vphi\Om\;,\quad \quad \nb\cdot\vec D = D_n\;\quad \mbox{ on } \pd_E\Om\;.
\end{split}
\end{equation}
The parts $\pd_u\Om$ and $\pd_\vphi\Om$ are nonvanishing, therefore, the  admissibility sets $\Ub(\Om)$, $\Vcal(\Om) $ introduced in Section~\ref{sec-Mac1} along with the associated spaces of test functions $\Ub_0(\Om)$, $\Vcal_0(\Om)$ are considered.
%
In the \emph{drained} case, pressure $\bar p$ is given. Consequently,  since equation \eq{eq-H11}$_3$ can be released, the formulation \eq{eq-H11} is reduced.
In the \emph{undrained} case, pressure $\bar p$ must be computed, thus, the formulation \eq{eq-H11} applies.
For the numerical tests reported below in Section~\ref{sec-num-ex}, $D_n=0$ and $\gb^s=0$.

\subsection{Model DFC*} 
In this case, part  $\pd_\vphi\Om$ vanishes, so that  \eq{eq-B2} is modified: we consider $\nb\cdot\vec D = D_n$ on the whole $\pd\Om$. The voltage $\bar\vphi^k$, $k=1,2$ is prescribed and the formulation \eq{eq-H11b*S} is employed.

\section{Reconstruction of the solutions at the microscopic level}\label{sec-reconstruct}

In this section we provide formulae which enable to reconstruct displacement,
pressure and velocity fields at the level of the heterogeneity's. This procedure
is affected by a given finite scale $\veps_0 >0$.

First we introduce the so-called folding procedure. The two-scale field
reconstruction is based on the coordinate split related to the periodic lattice.
For $\veps_0 > 0$, using the rescaled cell $Z^{\veps_0} = \veps_0 Y$ we
introduce its local copies $Z^{K,\veps_0}$ labeled by index $K$ whereby
$\{\bar{x}^K\}_K$ is the set of centers of each $Z^{K,\veps_0}$. For the sake
of simplicity, we consider only such domains $\Om$ which are generated as a
union of non overlapping RVEs $Z^{K,\veps_0}$, thus (recall that $\ol{Z}$ is the
closure of $Z$)
\begin{equation}\label{eq:Omega}
\ol{\Om} = \ol{\bigcup}_{K \in \Xi_\Om^{\veps_0}}  Z^{K,\veps_0}\;,\quad
Z^{K,\veps_0} = Z^{\veps_0} + \xibf^K\;,
\end{equation}
where $\Xi_\Om^{\veps_0}$ is the set of indices $K$ associated to the lattice
vector $\kb = (k_i) \in \ZZ^3$ such that $\xibf^K = \veps_0 k_i a_i$, recalling
the definition $Y = \prod_i ]-a_i/2,a_i/2[$. 

For any global position $x \in Z^{K,\veps_0}$, the local ``mesoscopic''
coordinate
\begin{equation}\label{eq:xy}
y = (x - \bar{x}^K)/\veps_0\;,
\end{equation}
can be introduced, such that $y \in Y$.  Then any two-scale function $f(x,y)$
can be evaluated by combining the macroscopic responses, such as displacements
$\ub^0(x)$, $x \in \Om$, and by the local ``autonomous'' characteristic
responses. Although, at
this point, we must distinguish between the two models described in the
preceding sections, the folding procedure can be summarized, as follows: for
each ``real sized'' cell $Y^{K,\veps_0}$ with its center $\bar{x}^K$  evaluate
the local responses given below as two-scale functions $f(x,y)$,  where $x \in
Z^{K,\veps_0}$ and $y \in Y$ is given by \eq{eq:xy}.

\subsection{Reconstruction for the CFD* model}

As the basis for reconstruction of the microscopic strains and the electric
field, the convergence result \eq{eq-CFD*2} and the split \eq{eq-L2}  of the
fluctuating fields $\ub^1$, $\vphi^1$ must be interpreted for a given size of
the microstructure, thus, for a given $\veps_0 >0$. This allows us to evaluate
approximations of $\ub^{R\veps_0}$ and $\vphi^{R\veps_0}$ by virtue of
\eq{eq-p3}. When dealing with the numerical implementation, an interpolation of
the macroscopic fields $\eebx{\ub^0}$ and $\nabla_x \vphi^0$  must be used to
introduce continuous two-scale functions  $f(x,y)$ at the global level for $x
\in \Om$, and $y \in Y$ associated with $x$ by \eq{eq:xy}. In particular, we can
introduce the Q1 interpolation scheme of the finite element method with
interpretation of the lattice formed by copies $Z^{K,\veps_0}$; let
$\Zcal_{\veps_0}(\Om)$ be such ``element partitioning''. By $\wtilde{g}$ we
denote the projection of a given function $g(x)$ in the space of piecewise Q1
polynomials defined over $\Zcal_{\veps_0}(\Om)$.

Now, recalling \eq{eq-p3},  we can express the reconstructed fields
$\ub^{R\veps_0}, \vphi^{R\veps_0}$ using the approximated two-scale functions,
\begin{equation}\label{eq-R0}
\begin{split}
\ub^{1,\veps_0}(x,y) & := \omegabf^{ij}(y)\wtilde{e_{ij}^x(\ub^0)} + \omegabf^k(y) \wtilde{\pd_k^x \vphi^0} - \bar p^0 \omegabf^P(y)\;,\\
\vphi^{1,\veps_0}(x,y) & := \eta^{ij}(y)\wtilde{e_{ij}^x(\ub^0)}+ \eta^k(y) \wtilde{\pd_k^x \vphi^0} - \bar p^0 \eta^P(y)\;,
\end{split}
\end{equation}
where $\bar p^0(x) = \bar p$ is the macroscopic pressure. When dealing with
disconnected inclusions, $\bar p^0(x)$ corresponds to the pressure in
$Z^{K,\veps}$ of the particular local RVE. Consequently, the convergence result
on the gradients yields the strain and the electric fields approximations,
\begin{equation}\label{eq-R1}
\begin{split}
\eb^\mic(x,y) = & \wtilde{\eebx{\ub^0}} + \eeby{\ub^{1,\veps_0}}\;,\\
\nabla \vphi^\mic \equiv \vec E^\mic(x,y) = & \wtilde{\nabla_x \vphi^0} + \nabla_y \vphi^{1,\veps_0}\;,
\end{split}
\end{equation}
which can be rewritten in terms of the local response gradients:
\begin{equation}\label{eq-R2}
\begin{split}
\eb^\mic(x,y) = & \wtilde{\eebx{\ub^0}} + \eeby{\omegabf^{ij}}\wtilde{e_{ij}^x(\ub^0)}
+ \eeby{\omegabf^{k}}\wtilde{\pd_k^x \vphi^0} - \eeby{\omegabf^P}\bar p^0\;,\\
\vec E^\mic(x,y) = & \wtilde{\nabla_x \vphi^0} + \nabla_y \eta^{ij}\wtilde{e_{ij}^x(\ub^0)}
+ \nabla_y \eta^k\wtilde{\pd_k^x \vphi^0}  - \nabla_y \eta^P \bar p^0\;.
\end{split}
\end{equation}
It should be noted that  \eq{eq-R1}$_1$ and \eq{eq-R1}$_2$ can be extended
naturally by terms $\veps_0 \eebx{\ub^{1,\veps_0}}$ and $\veps_0
\nabla_x{\vphi^{1,\veps_0}}$, respectively.

\subsection{Reconstruction for the DFC* model}

In this case, the reconstruction of the microscopic strains and the electric
field follows the analogous guidelines as those explained above for the CFD*
model. Therefore, we only describe the major differences. Importantly, the
computational material parameters $\bar \parg$ and $\bar \db$ must be defined for
a given $\veps_0$ using \eq{eq-S*1}, where $\parg^{\veps_0}$ and $\db^{\veps_0}$
present the right physical values.

For the reconstruction of displacements $\ub^{R\veps_0}$ and $\vphi^{R\veps_0}$,
in \eq{eq-S*3}, the relevant two-scale functions are expressed according to the
decomposed form \eq{eq-L2b} modified due to the projection of the macroscopic
strains into the Q1 polynomial bases associated with the partitioning
$\Zcal_{\veps_0}(\Om)$ introduced in the preceding section. Thus, since
$\vrho_E$ and $\bar \vphi^k$ are assumed continuous (or even constant) functions
in $x$, \eq{eq-L2b} yields:
\begin{equation}\label{eq-R3}
\begin{split}
\ub^{1,\veps_0}(x,y) & = \omegabf^{ij}(y)\wtilde{e_{ij}^x(\ub^0)}  - p^0(x) \omegabf^P(y) + \omegabf^\rho(y) \rho_E + \sum_k\hat\omegabf^k(y)\bar\vphi^k\;,\\
\vphi^{0,\veps_0}(x,y) & = \hat\eta^{ij}(y)\wtilde{e_{ij}^x(\ub^0)} - p^0(x) \hat\eta^P(y) + \hat\eta^\rho(y) \rho_E + \sum_k\hat\vphi^k(y)\bar\vphi^k\;.
\end{split}
\end{equation}
Then the convergence result on the gradients, see \eq{eq-S*5}$_{2,4}$ yields the
strain and the electric fields approximations,
\begin{equation}\label{eq-R4}
\begin{split}
\eb^\mic(x,y) = & \wtilde{\eebx{\ub^0}} + \eeby{\ub^{1,\veps_0}}\;,\\
\nabla \vphi^\mic \equiv \vec E^\mic(x,y) = & \frac{1}{\veps_0}  \nabla_y \vphi^{0,\veps_0}\;,
\end{split}
\end{equation}
where the substitutions due to \eq{eq-R3} are obvious. Beyond the first order
approximation, \eq{eq-R4}$_1$ and \eq{eq-R4}$_2$ can be extended naturally by
terms $\veps_0 \eebx{\ub^{1,\veps_0}}$ and $\nabla_x{\vphi^{0,\veps_0}}$,
respectively; note the difference with the CFD* model.

\section{Numerical examples}\label{sec-num-ex}

In this section we demonstrate two-scale modelling of the porous piezoelectric
media using the two homogenized models described above. The numerical
simulations presented in this section has been performed in {\it SfePy} --
Simple Finite Elements in Python, see \cite{sfepy}. It is a software for solving
multiscale systems of coupled partial differential equations by means of the
finite element method. Both the models, CFD* and DFC*, have been implemented
in this software, whereby the displacements and the electric potential were
approximated by piecewise three-linear functions, thus, the Q1 hexahedral
elements are employed. For the model DFC* with fluid inclusions the pressure  is
approximated by the piecewise constant functions.

\subsection{Validation test: homogenization vs. reference model}\label{sec:ex_valid}

To validate the proposed homogenized model of fluid-saturated piezoelectric
porous media, we rely on the numerical results obtained by direct simulations of
the heterogeneous periodic structure. Thus, the reference numerical model is
established by copies $Z^{K,\veps_0}$ of the reference cell $Z^{\veps_0}$, as
discussed in Section~\ref{sec-reconstruct}. To capture accurately effects
related to the microstructure geometry, highly refined finite element meshes
associated with all cells $Z^{K,\veps_0}$  are required for a given size
$\veps_0>0$. For the validation test we consider the homogenized model DFC*
featured by disconnected fluid inclusions and connected conductor fibers. The
local problems  \eq{eq-ch1S}--\eq{eq-ch5S} are solved in the unit cell $Y$
represented by the rescaled finite element mesh. Responses computed by the
macroscopic model  represented by  \eq{eq-H*S}, \eq{eq-H11b*S} are compared to
the responses of the poroelastic-piezoelectric reference model which is defined
by the equilibrium equations \eq{eq-2}, \eq{eq-3a}, and by interface and
boundary conditions \eq{eq-4a} and \eq{eq-4b}. In this section, we use
subscripts $x,y,z$  to refere to the coordinate axes directions, thus, we write
$u_y$ instead of $u_2$, etc.

\begin{figure}[ht]
\centering
\includegraphics[height=4cm]{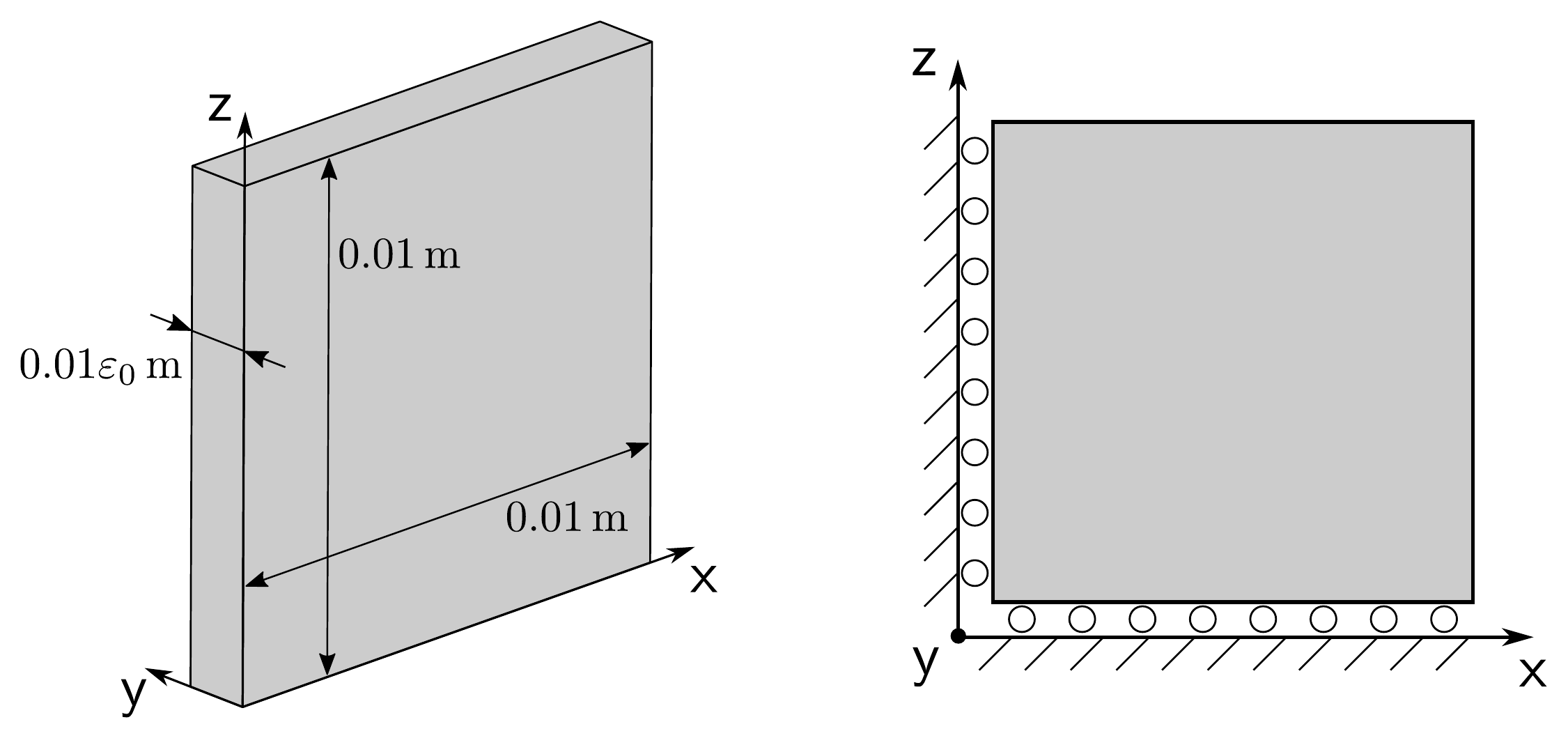}
\caption{Sample used in the validation test: left -- shape and dimensions of the
sample; right -- applied boundary conditions, $u_y = 0$ at the bottom face, $u_x =
0$ at the left face, periodic conditions in $y$ direction at the front and back
faces.}
\label{fig:ex_valid1}
\end{figure}

We consider a block sample of dimensions $0.01 \times 0.01\veps_0 \times
0.01$\,m on which we apply the following boundary conditions: $u_y = 0$ at the
bottom face, see Fig.\ref{fig:ex_valid1} right, $u_x = 0$ at the left face and
the periodic condition in $y$ direction (front and back faces). No volume and
surface forces and surface electric charge are considered, thus, $\hat\fb =
\bmi0$, $\bar\hb = \bmi0$, $\varrho_E = 0$ in \eq{eq-H11b*S}. The piezoeletric
matrix is made of barium--titanite BaTiO$_{3}$, see Tab.~\ref{tab:pz_material}
for its material properties, the properties of metallic conductors are given by
Young's modulus $E = 200$\,GPa and Poisson's ration $\nu = 0.25$, and the fluid
compressibility is $4.651 \times 10^{-10}$\,Pa.

\begin{table}[ht]
\begin{tabular}{lrrrrrr}
\hline
elasticity (in GPa): & $A_{1111}$ & $A_{3333}$ & $A_{1122}$ & $A_{2233}$ & $A_{1313}$ & $A_{1212}$ \\
      & 15.040 & 14.550 & 6.560 & 6.590 & 4.240 & 4.390 \\
\hline
\end{tabular}

\begin{tabular}{lrrrr}
piezo-coupling  (in C/m$^2$): & $g_{311}$ & $g_{322}$ & $g_{333}$ & $g_{223}$ \\
  & -4.322 & -4.322 & 17.360 & 11.404 \\
\hline
\end{tabular}

\begin{tabular}{lrr}
dielectricity (in $10^{-8}$ C/Vm ): & $d_{11}$ & $d_{33}$ \\
      & 1.284 & 1.505\\
\hline
\end{tabular}

\caption{Piezoelectric properties of the porous matrix. The transverse isotropy
yields the following symmetries: $A_{2233}=A_{1133}$, $A_{1313} = A_{2323}$,
$g_{311}= g_{322}$, $g_{223}= g_{113}$, $d_{11} = d_{22}$. Other components are
zero.}\label{tab:pz_material}

\end{table}

\begin{figure}[ht]
\centering
\includegraphics[height=9cm]{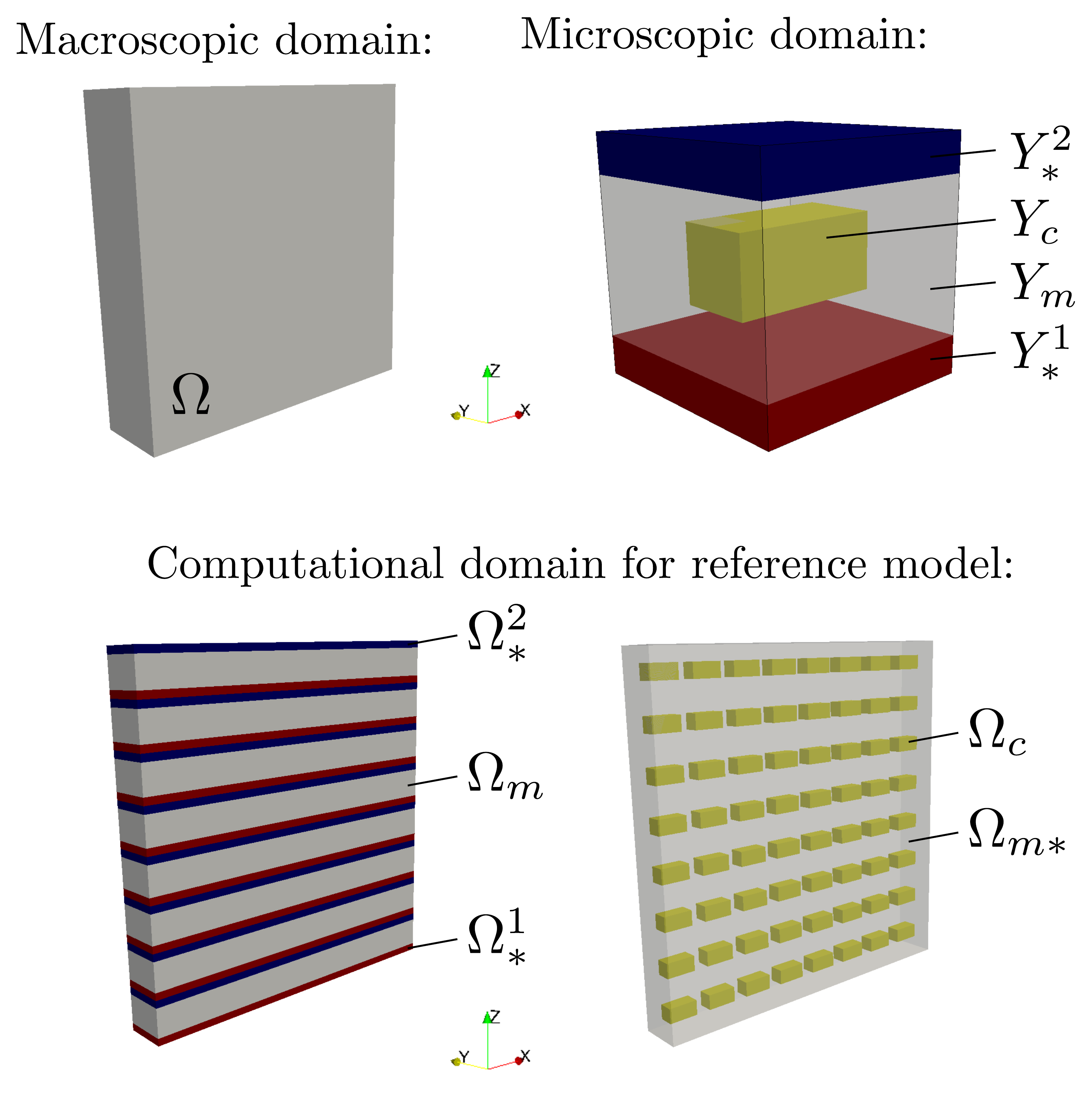}
\caption{Computational domains: top -- macroscopic domain $\Omega$ (left) and
decomposition of microscopic domain $Y$ (right); bottom -- domain use in the
reference model build up by repetition of the microscopic unit.}
\label{fig:ex_valid2}
\end{figure}

The homogenized response is obtained by solving separately the  macroscopic and
the microscopic problems;  the corresponding domains $\Om$ and $Y$ for these subproblems
are depicted in Fig.~\ref{fig:ex_valid2} top. The finite element mesh used in
the reference model is build up as an array of cells $Z^{K,\veps_0}$, by
repeting the representative cell $Z^{\veps_0}$ in the $x$ and the
$z$ directions, see Fig.~\ref{fig:ex_valid2} top-right and bottom. The number of
copies in these two directions is referred as a ``grid`` in the subsequent
text and figures. There are no external loads prescribed, so that the deformation of the sample is induced due to the
piezoelectric effect, as the response to the locally prescribed potentials $\bar\vphi^1 = +1000$\,V
and $\bar\vphi^2 = -1000$\,V associated with the two condustors. In the case of the
reference model, these potentials are applied as the interface conditions, see
\eq{eq-4a}, while in the homogenized model they appear at the \rhs of
\eq{eq-H11b*S}.

\begin{figure}[ht]
\centering
\includegraphics[height=14cm]{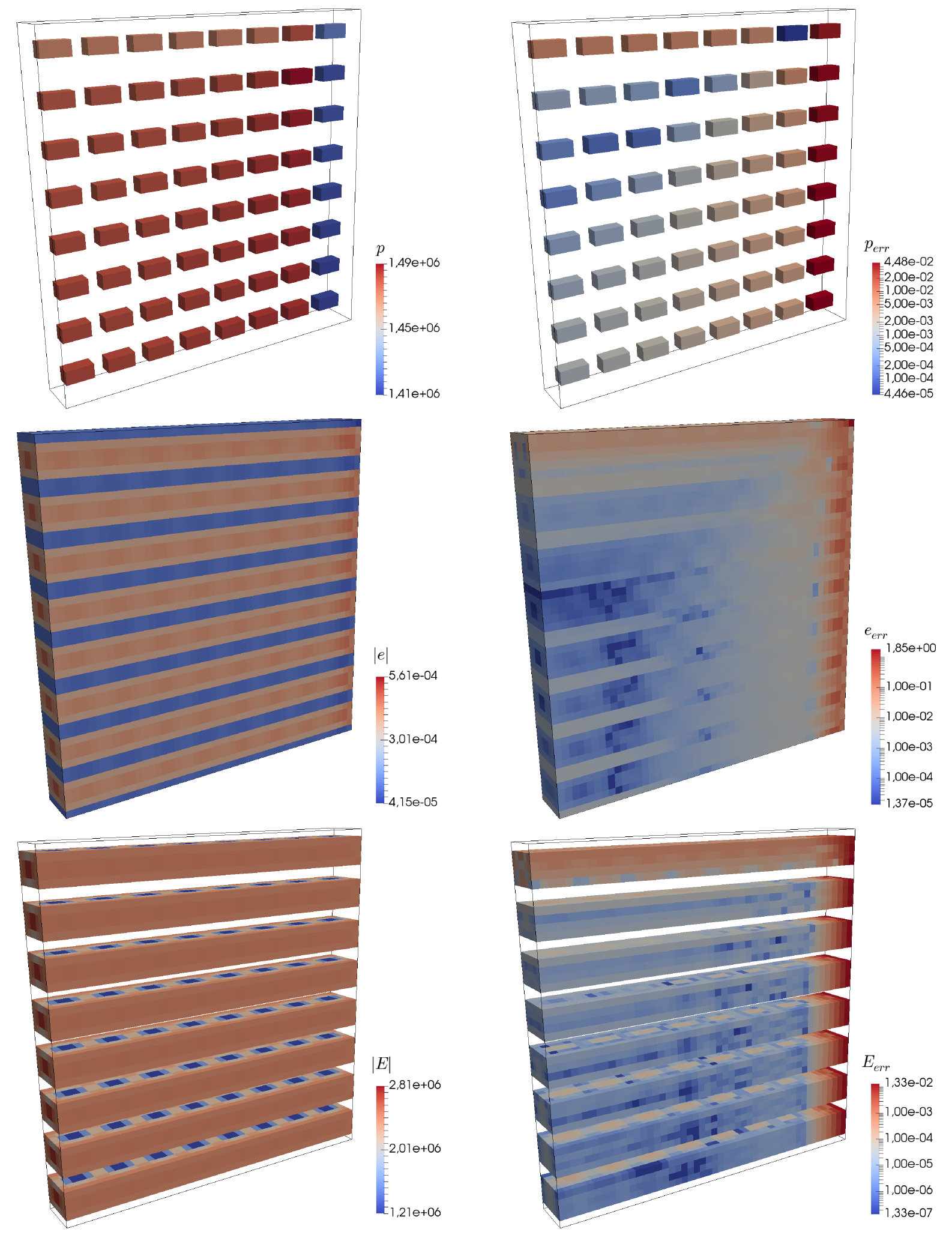}
\caption{Left -- responses of the reference model: fluid pressure $p$, strain
field magnitude $\norm{\eb}$, electric field magnitude $\norm{\vec E}$; right --
relative errors of the results obtained by the homogenized model.}
\label{fig:ex_valid3}
\end{figure}

\begin{figure}[ht]
\centering
\includegraphics[height=9cm]{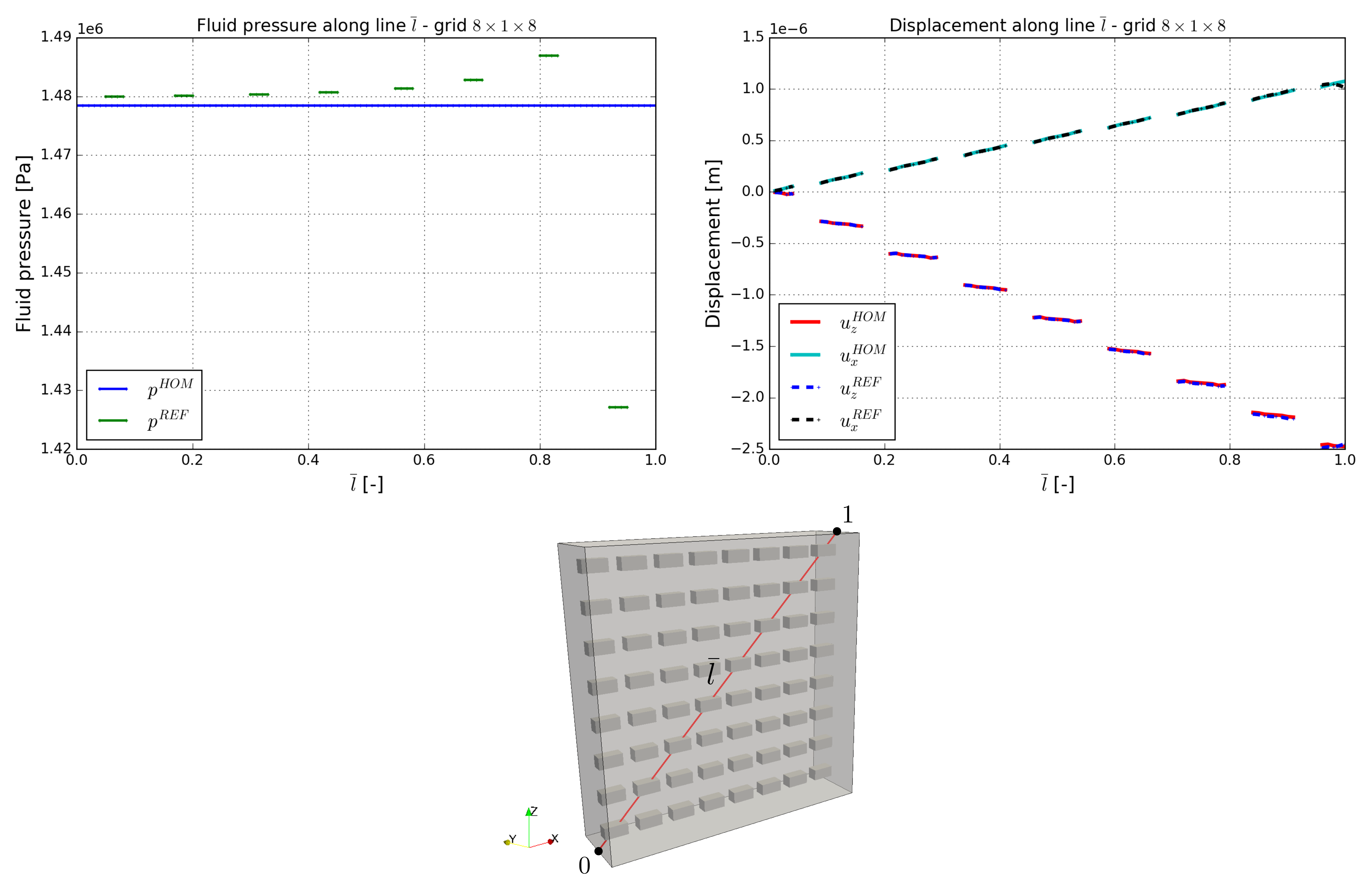}
\caption{Responses of the homogenized and reference model over line
$\bar l$: left -- fluid pressure; right -- displacements $u_x$, $u_z$.}
\label{fig:ex_valid3b}
\end{figure}

\begin{figure}[ht]
\centering
\includegraphics[height=4cm]{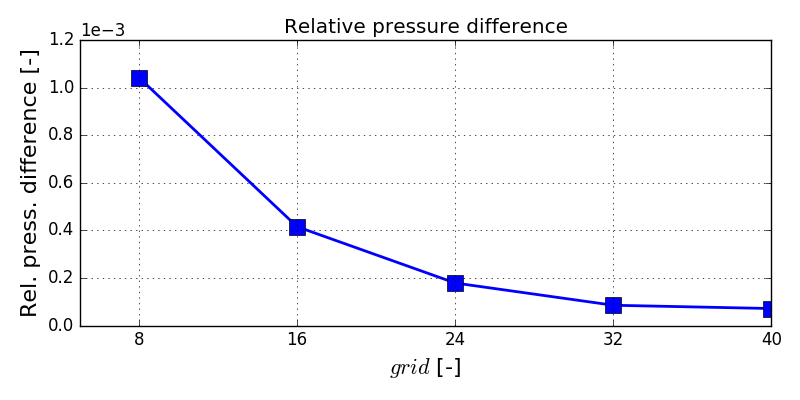}
\caption{Drop of the relative pressure error with the increasing grid number.}
\label{fig:ex_valid4}
\end{figure}

The responses of the reference model are shown in Fig.~\ref{fig:ex_valid3} left,
where the fluid pressure in the inclusions, strain field and electric field are
depicted. The right part of Fig.~\ref{fig:ex_valid3} shows the relative error of
the fields calculated by the homogenized model. The strain and electric fields
are reconstructed according to \eq{eq-R3}, \eq{eq-R4}, whereby  the homogenized
fluid pressure results from macroscopic equations \eq{eq-H11b*S}. The relative
error for the pressure, strains, and  the electric field displayed in the figure
is defined as $f_{err} = {\vert\, f^{dir} - f^{hom} \vert / f^{dir}}$, where $f$
stands for $p$, $\norm{\eb}$ and $\norm{\vec E}$, respectively. In
Fig.~\ref{fig:ex_valid3b}, fluid pressures and displacements $u_x$, $u_z$ along
line $\bar l$ are compared.
The biggest differences apparent near the right and top borders of the
sample appear there because the periodicity assumption applied in the
homogenized model is not satisfied at this boundary. When moving to the left
bottom part, where symmetric boundary conditions are applied, the boundary effect
vanishes and the relative difference  drops significantly.
Figure \ref{fig:ex_valid4} shows how the relative pressure error is decreasing
with the increasing grid number, i.e. $\veps_0 \longrightarrow 0$.

For the reference model, the solution time of the validation test
with $\veps_0 = 0.01 /$\,grid, where \hbox{grid = 24}, 
is about 300~seconds and the finite element model has approximately $4.5 \times
10^{5}$ degrees of freedom (DOFs). For the corresponding homogenized medium, the
solution time is about 20~seconds including reconstructions of the fields at a
given macroscopic element, which take the most amount of the computational time.
The microscopic problems with 741 DOFs are solved several times to compute all the
corrector functions, then the macroscopic problem with 577~DOFs is solved.

\subsection{DFC* example}

\begin{figure}[ht]
\centering
\includegraphics[height=4.2cm]{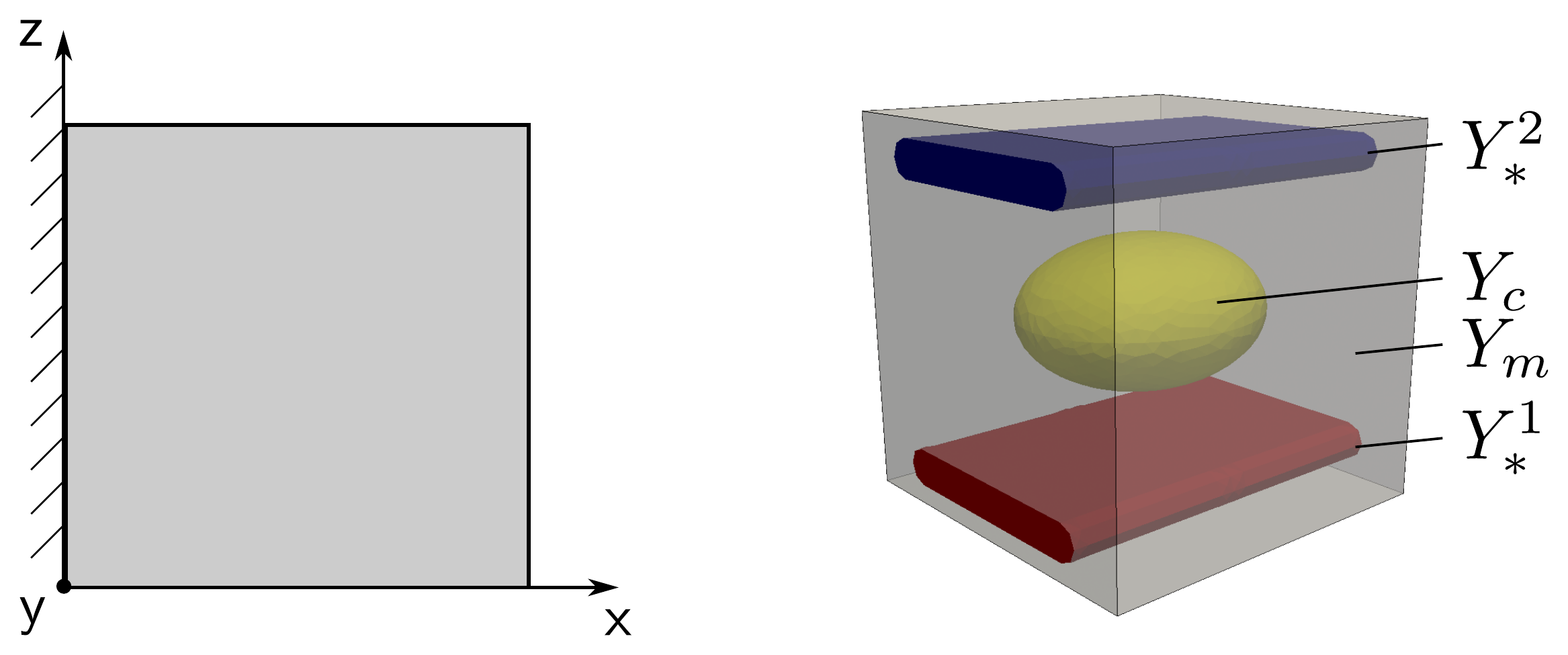}
\caption{Left -- boundary conditions applied at the macroscopic level; right --
geometry of the DFC* microscopic periodic cell (RVE).}
\label{fig:ex_dfc1}
\end{figure}

\begin{figure}[ht]
\centering
\includegraphics[height=4cm]{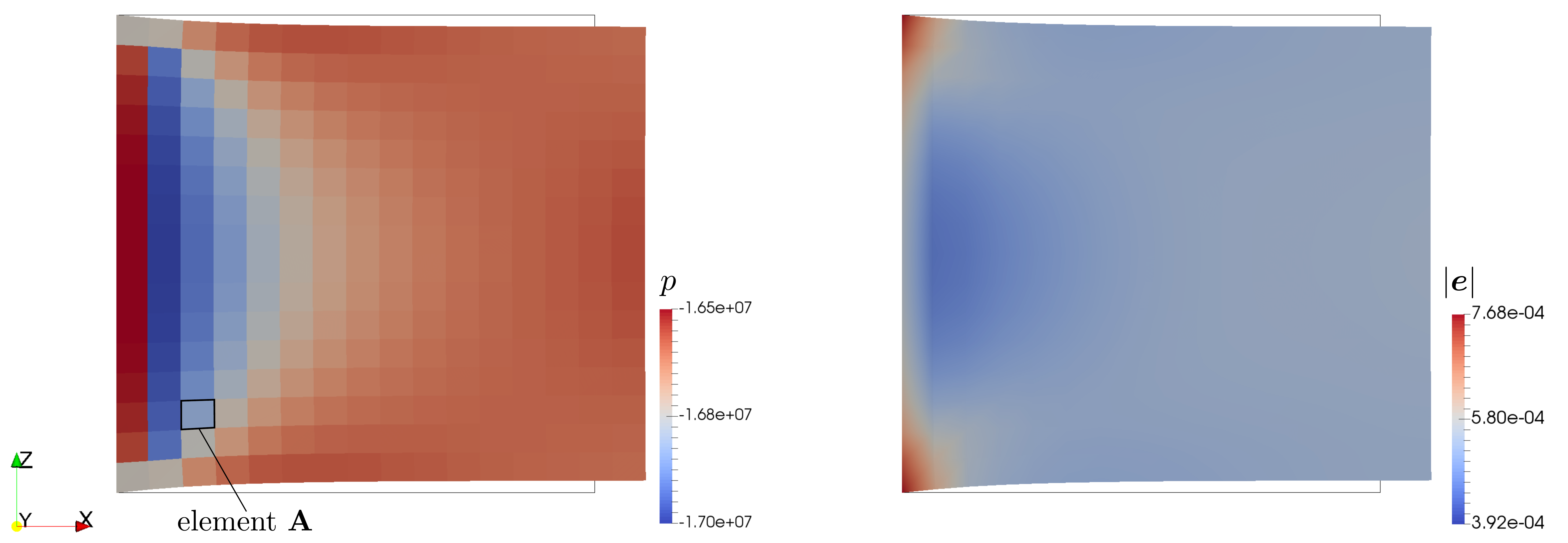}
\caption{Deformed macroscopic (displacements scaled by factor 300) sample and
the resulting fields, DFC* model: left -- pressure $p$; right -- magnitude of
macroscopic strain $\eb(\ub^0)$.}
\label{fig:ex_dfc2}
\end{figure}

\begin{figure}[ht]
\centering
\includegraphics[height=4.8cm]{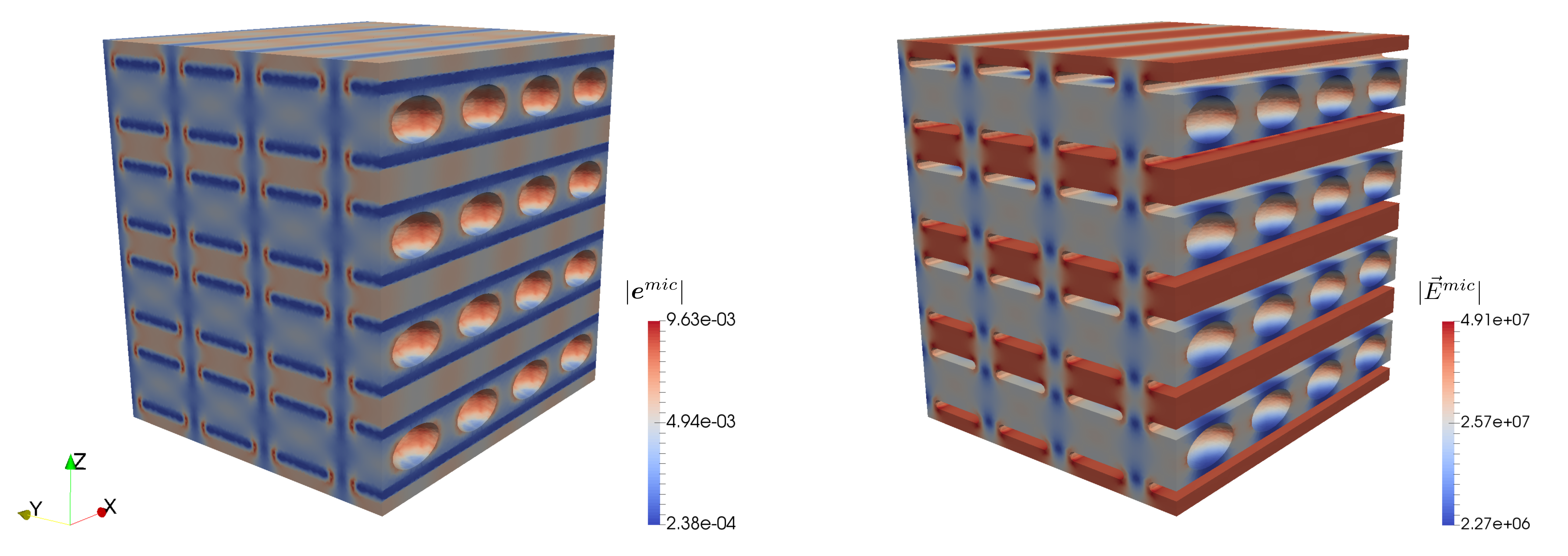}
\caption{Magnitudes of reconstructed fields at the macroscopic element {\bf A}
(see Fig.~\ref{fig:ex_dfc2}), DFC* model: left -- strain field $\eb^{mic}$;
right -- electric field $\vec E^{mic}$.}
\label{fig:ex_dfc3}
\end{figure}

As in the preceding section, we consider the isolated fluid inclusions and two
continuous conductors embedded in the piezoelectric matrix, see
Fig.~\ref{fig:ex_dfc1} right. The left face of the macroscopic sample, with
dimensions $0.01 \times 0.0025 \times 0.01$\,m, is attached to the rigid wall
, see Fig.~\ref{fig:ex_dfc1} left, the periodic boundary condition is applied in $y$
direction and the deformation is induced by prescribing potentials $\pm$~1000\,V
in the embedded conductors. The deformed shape (deformation scaled by factor
300) of the sample and pressure field $p$ and the magnitude of macroscopic
strain $\eb(\ub^0)$ are depicted in Fig.~\ref{fig:ex_dfc2}. The reconstructed
strain and electric fields for  $\veps_0 = 0.01 / 64$ (grid = 64) at a
given macroscopic element are shown in Fig.~\ref{fig:ex_dfc3}.

\subsection{CFD* example}

\begin{figure}[ht]
\centering
\includegraphics[height=4.2cm]{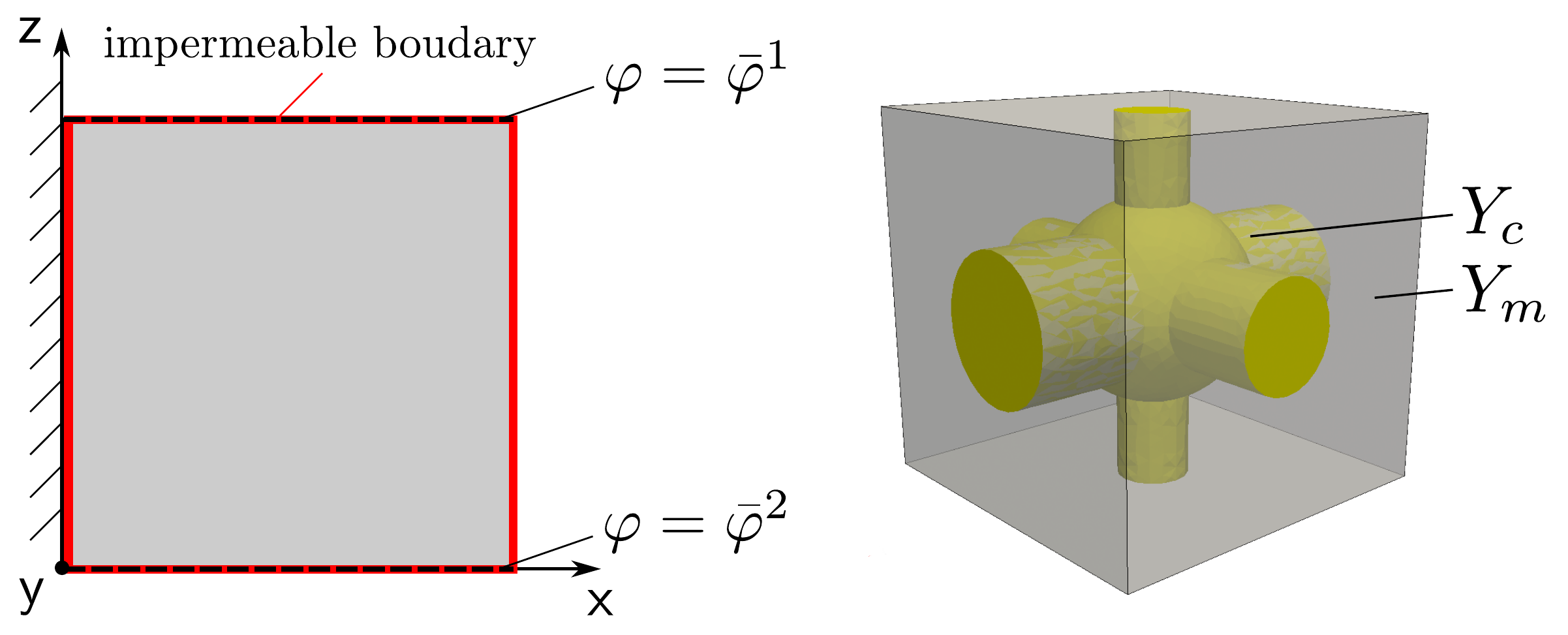}
\caption{Left -- boundary conditions applied at the macroscopic level; right --
geometry of the CF(D*) microscopic periodic cell (RVE).}
\label{fig:ex_cf1}
\end{figure}

In this part, we present the numerical simulation of the piezoelectric medium
with connected fluid channels governed by equations \eq{eq-ch1}--\eq{eq-ch3} and
\eq{eq-H5a}--\eq{eq-H11}. For the sake of simplicity the structure without
conductor inclusions is considered. We omit them because they are passive
elements only, contrary to the previous examples. The sample is again
fixed on its left face, the periodic condition is applied in $y$ direction and
the deformation is now invoked by the prescribed electrical potentials at the
bottom and top face of the sample, see Fig.~\ref{fig:ex_cf1} left,
$\bar\vphi^1 = -1000$\,V, $\bar\vphi^2 = +1000$. The fluid pressure, $\bar
p$ in \eq{eq-H11}, is constant in the whole macroscopic domain and can be
treated either as a unknown variable in the case when the fluid pores are closed
on the outer surface (impermeable boundary) -- undrained case, or as a given
value determined by the pressure of a surrounding medium -- drained case.

\begin{figure}[ht]
\centering
\includegraphics[height=4cm]{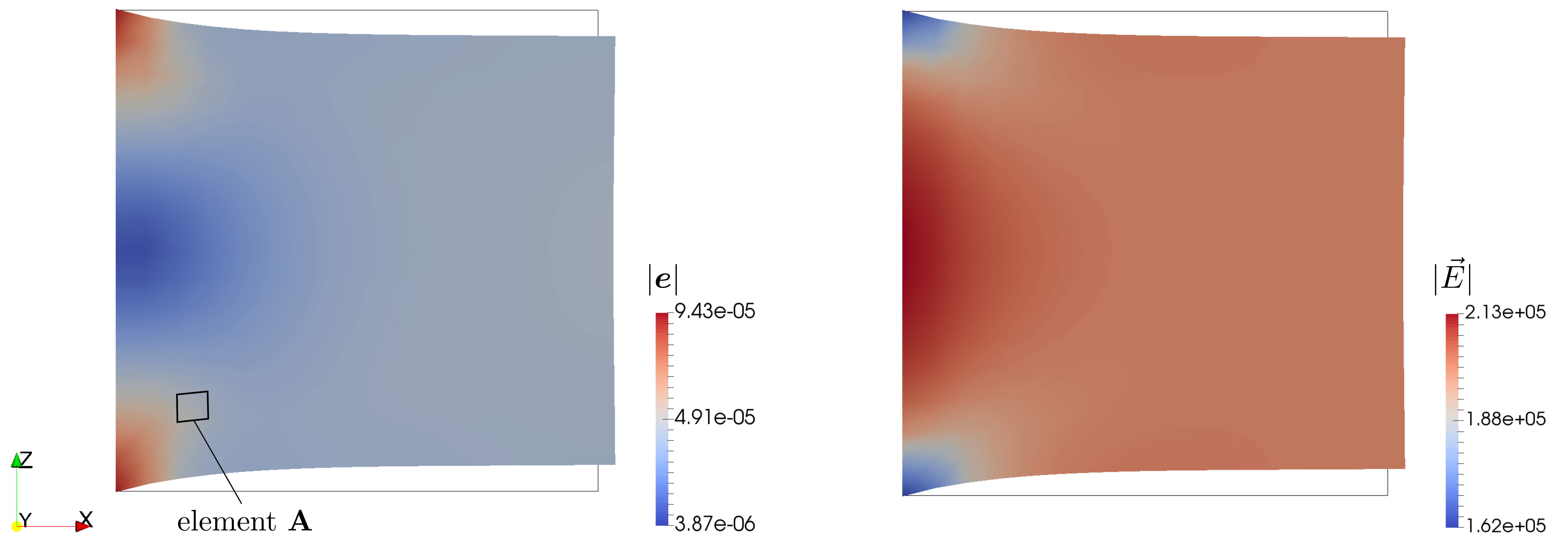}
\caption{Deformed macroscopic (displacement scaled by factor 3000) sample and
the resulting magnitudes of the macroscopic fields, CFD* model: left -- strain
$\eb(\ub^0)$; right -- macroscopic electric field $\vec E = \nabla_x \varphi^0$.}
\label{fig:ex_cf2}
\end{figure}

\begin{figure}[ht]
\centering
\includegraphics[height=4.8cm]{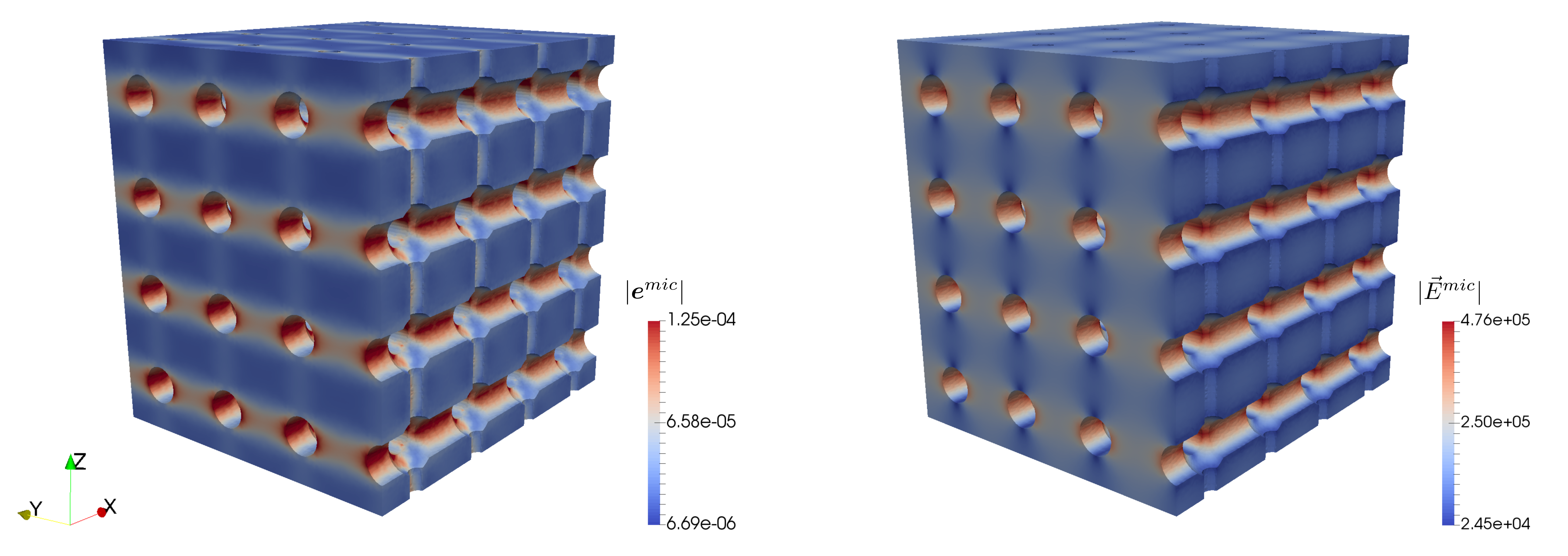}
\caption{Magnitudes of reconstructed fields at the macroscopic element {\bf A}
(see Fig.~\ref{fig:ex_cf2}), CFD* model: left -- strain field $\eb^{mic}$;
right -- electric field $\vec E^{mic}$.}
\label{fig:ex_cf3}
\end{figure}

The macroscopic responses and reconstructed strain and electric fields for the
undrained case are depicted in Figs.~\ref{fig:ex_cf2}, \ref{fig:ex_cf3}. The
computed homogenized coefficients $\Bb^H$, $M^H$, $\coefF$ and $\Aop^H$,
$\coefG$, $\Db^H$ are summarized in Tables~\ref{tab:ex_cf1} and
\ref{tab:ex_cf2}, where $\Aop^H$, $\coefG$, $\Db^H$ are compared to the
material properties of the homogeneous solid skeleton. The finite size of the
microstructure is given by $\veps_0 = 0.01 / 64$.

\begin{table}[ht]
\begin{tabular}{lcccc}
  \hline
 & $B^H_{11}$ & $B^H_{33}$ & $M^H$ & $F^H_3$ \vbox to 1.2em{}\\
 & 0.347 & 0.370 & 4.834 $\times 10^{-121}$\,Pa & -4.464 $\times 10^{-12}$\, m/V \\
\hline
\end{tabular}
\caption{Homogenized coefficients $\Bb^H$, $M^H$, $\coefF$.}\label{tab:ex_cf1}
\end{table}

\begin{table}[ht]
\begin{tabular}{lrrrrrr}
\hline
elasticity (in $10^{10}$ Pa):\vbox to 1.2em{} & $A^H_{1111}$ & $A^H_{3333}$ & $A^H_{1122}$ & $A^H_{2233}$ & $A^H_{1313}$ & $A^H_{1212}$ \\
homogen. prop.      & 9.745 & 7.888 & 2.582 & 2.376 &  3.452 & 3.114\\
skeleton prop.      & 15.040 & 14.550 & 6.560 & 6.590 & 4.240 & 4.390 \\
\hline
\end{tabular}

\begin{tabular}{lcccc}
piezo-coupling (in C/m$^2$):\vbox to 1.2em{} & $G^{H}_{311}$ & $G^{H}_{322}$ & $G^{H}_{333}$ & $G^{H*}_{223}$ \\
homogen. prop.  & -1.484 & -1.142 & 11.348 & 5.761 \\
skeleton prop.  & -4.322 & -4.322 & 17.360 & 11.404 \\
\hline
\end{tabular}

\begin{tabular}{lcc}
dielectricity (in $10^{-8}$ C/Vm ):\vbox to 1.2em{} & $D^{H}_{11}$ & $D^{H}_{33}$ \\
homogen. prop.      & 0.997 & 0.988\\
skeleton prop.      & 1.284 & 1.505 \\
\hline
\end{tabular}
\caption{Homogenized piezoelectric coefficients $\Aop^H$, $\coefG$,
$\Db^{H}$ compared to the properties of the homogeneous piezoelectric
material.}\label{tab:ex_cf2}
\end{table}

\section{Conclusion}

We considered heterogeneous periodic microstructure consisting of a
piezoelectric skeleton with embedded conducting parts and penetrated by channels
filled with a electrically neutral fluid. Using the the homogenization method we
derived macroscopic models of the fluid saturated porous piezoelectric material.
We elaborated two models for different topologies of the pores and arrangements
of metallic conducting parts.

The model CFD* is characterized by a single connected porosity, whereas the
metal parts can be distributed as small inclusions. The obtained effective
constitutive law for the stress and the electric displacement involves new
coefficients related to the pore fluid pressure. As the consequence, by
increasing the fluid pressure, or the pore fluid volume, the electric field can
be generated, whereby the inverse effect appears, being naturally consistent
with Onsager reciprocity principles. The macroscopic model derived by upscaling
from the level of heterogeneities is consistent with the phenomenological
models, \cf \cite{Sharma-PRSA2010-waves-pz,Vashishth-wave-pz-2009}.

The periodic structure for model DFC* is  characterized by fluid inclusions
embedded in the piezoelectric matrix and by two, or more metallic electrodes
being  embedded in the matrix. In the latter case, an external electric field is
imposed through given electric potentials associated with each of the
electrodes. This electric field intensity blows up with $\veps^{-1}$ when
$\veps\rightarrow 0$, \ie  decreasing the microstructure characteristic size. To
compensate the effect, a weakly piezoelectric material must be considered;  the
coupling and the dielectricity constants of the skeleton piezoelectric material
are scaled proportionally to the microstructure size $\veps$, so that the limit
macroscopic model could be obtained. The obtained model describes a metamaterial
with an interesting property of locally controllable pore fluid pressure. This
option will be pursued further to develop a model of  fluid transport
controllable by voltage distributed by means of electrodes penetrating into the
periodic medium structure.

For both the models we presented the microscopic level response reconstruction
which is based on the characteristic responses and the macroscopic solutions of
the particular problem. This was employed to validate the homogenized medium
models. As the reference model for the validation, we used the direct numerical
finite element simulation of a given periodic heterogeneous piezoelectric
material interacting with the fluid. 

\paragraph{Acknowledgment} This research is supported by project GACR 16-03823S
and in part by project LO 1506 of the Czech Ministry of Education, Youth and
Sports.

\section*{Appendix A}\label{sec-appendix}
We prove the symmetry relationships \eq{eq-H8}.
The first equality \eq{eq-H8}$_1$ follows from \eq{eq-ch3} and \eq{eq-ch2},
\begin{equation}\label{eq-H9}
\begin{split}
& \intY_{Y_m}\nabla_y\cdot\omegabf^i\dY = \aYms{\omegabf^P}{\omegabf^i}-\gYm{\omegabf^i}{\eta^P} \\
& =  \gYm{\omegabf^P}{\eta^i+y_i} + \dYm{\eta^i+y_i}{\eta^P} = \gYm{\omegabf^P}{y_i} + \dYm{y_i}{\eta^P}\;.
\end{split}
\end{equation}
To show \eq{eq-H8}$_2$, we employ \eq{eq-ch1} and \eq{eq-ch2}, which yields:
\begin{equation}\label{eq-H9a}
\begin{split}
-\aYms{\omegabf^k}{\Pibf^{ij}} & = \aYms{\omegabf^{ij}}{\omegabf^k} - \gYm{\omegabf^k}{\eta^{ij}}\;,\\
\gYm{\Pibf^{ij}}{\eta^k} & = \dYm{\eta^{ij}}{\eta^k} - \gYm{\omegabf^{ij}}{\eta^k} \\
& = \dYm{\eta^{ij}}{y_k} + \gYm{\omegabf^k}{\eta^{ij}} + \gYm{\omegabf^{ij}}{y_k} - \aYms{\omegabf^k}{\omegabf^{ij}}\;.
\end{split}
\end{equation}
Upon summation the above two equalities the equality \eq{eq-H8}$_2$ follows.
In analogy, to show \eq{eq-H8}$_3$, we employ \eq{eq-ch1} and \eq{eq-ch3}; the following equalities
\begin{equation}\label{eq-H9b}
\begin{split}
\aYms{\omegabf^P}{\Pibf^{ij}} & = \aYms{\omegabf^{ij}}{\omegabf^P} - \gYm{\omegabf^P}{\eta^{ij}} \\
& = \dYm{\eta^P}{\eta^{ij}} +  \gYm{\omegabf^{ij}}{\eta^P} + \intY_{\Gamma_Y}\omegabf^{ij}\cdot \nb^\mx\dSy\;,\\
\gYm{\Pibf^{ij}}{\eta^P} & = -\dYm{\eta^{ij}}{\eta^P} - \gYm{\omegabf^{ij}}{\eta^P}\;.
\end{split}
\end{equation}
Upon summation the above two equalities the equality \eq{eq-H8}$_3$ follows.

The symmetric expressions of $\Aop^H = (A_{ijkl}^H )$ and $\Db^H = (D_{kl}^H)$ are obtained due to the local problems \eq{eq-ch1} and \eq{eq-ch2}.
Obviously, the tensors $\Aop^H = (A_{ijkl}^H )$, and $\Bb^H = (B_{ij}^H )$ are symmetric, $\Aop^H$ inherits all the symmetries of
$\Aop$; moreover $\Aop$ is positive definite and $M^H > 0$.





\end{document}